\catcode`\@=11

\expandafter\ifx\csname draftcontrol\endcsname\relax\global\def\draftcontrol{0}
\fi

{\count255=\time\divide\count255 by 60
\xdef\hourmin{\number\count255}
\multiply\count255 by-60\advance\count255 by\time
\xdef\hourmin{\hourmin:\ifnum\count255<10 0\fi\the\count255}}
\def\draftdate{\number\month/\number\day/\number\year\ \ \ \hourmin }

\newcommand\makepapertitle{\par
  \begingroup
    \renewcommand\thefootnote{\@fnsymbol\c@footnote}%
    \def\@makefnmark{\rlap{\@textsuperscript{\normalfont\@thefnmark}}}%
    \long\def\@makefntext##1{\parindent 1em\noindent
            \hb@xt@1.8em{%
                \hss\@textsuperscript{\normalfont\@thefnmark}}##1}%
     \newpage
     \global\@topnum\z@   % Prevents figures from going at top of page.
     \@makepapertitle
     \thispagestyle{empty}\@thanks
  \endgroup
  \setcounter{footnote}{0}%
  \global\let\thanks\relax
  \global\let\makepapertitle\relax
  \global\let\@makepapertitle\relax
  \global\let\@thanks\@empty
  \global\let\@author\@empty
  \global\let\@date\@empty
  \global\let\@title\@empty
  \global\let\title\relax
  \global\let\author\relax
  \global\let\date\relax
  \global\let\and\relax
  \def\version{\let\version\@version\@gobble}
}
\def\@makepapertitle{%
  \newpage
   \ifnum\draftcontrol=1 {}
   \version\versionno
   \vskip 3em%
   \else
   \hfill\hbox to 3cm {\parbox{4cm}{\@pubnum}\hss}%
   \vskip 3em%
   \fi
   \begin{center}%
   \let \footnote \thanks
     {\LARGE \@title \par}%
     \vskip 1.5em%
     {\normalsize%\large
       \lineskip .5em%
       \begin{center} %\begin{tabular}[t]{c}%
         \@author
       \end{center} % \end{tabular}
\par}%
     \vskip 1em%
     {\@bstract}%
     \end{center}%
     \vskip .5em
     \@date%
   \par
}

\gdef\@pubnum{}
\def\pubnum#1{%
  \gdef\@pubnum{#1}}

\gdef\@bstract{}
\def\Abstract#1{%
  \gdef\@bstract{%
   \parbox{\textwidth-0pc}{%
   \centerline{\bf Abstract}\penalty1000
   \noindent%\abstractfont \baselineskip=12pt
   \renewcommand\baselinestretch{1.0}
   {#1}}}
}

\def\ps@paper{\let\@mkboth\@gobbletwo%
     \ifnum\draftcontrol=1
        \def\@oddfoot{\hbox to \textwidth{\tiny \versionno \hfil\tiny\draftdate}%
        \hskip -\textwidth \hbox to \textwidth{\hfil\rm\thepage\hfil}}%
     \else\def\@oddfoot{\hbox to \textwidth{\hfil\rm\thepage\hfil}}
     \fi
     \let\@evenfoot\@oddfoot
}
\def\body{\clearpage
          \pagestyle{paper}
        }
\def\@version#1{\ifnum\draftcontrol=1
\typeout{}\typeout{#1}\typeout{}
\vskip3mm\centerline{\hbox{\fbox{\normalsize{\tt DRAFT -- #1 -- }
                   {\draftdate}}}}\vskip3mm
\fi}
\let\version\@version
\long\def\eqlabel#1{\ifnum\draftcontrol=1
                    \tag@false  % there are some problems with multline without this
                    \tag*{(\theequation) \hbox to -0.2cm{\hspace{0cm}\small{#1}\hss}}
                    \refstepcounter{equation} 
                    \edef\@currentlabel{\theequation}
                    \ltx@label{#1}          % use old LaTeX \label instead of new definition
                    \else
                    \label{#1}
                    \fi
                    }
\let\st@bibitem\@bibitem
\let\st@lbibitem\@lbibitem
\ifnum\draftcontrol=1
  \def\@bibitem#1{%
    \st@bibitem{#1}\a@@label{#1}\ignorespaces}
  \def\@lbibitem[#1]#2{%
    \st@lbibitem[#1]{#2}\a@@label{#2}\ignorespaces}
  \def\a@@label#1{%
    \gdef\a@lab{\smash{\normalfont\small#1}}
    \ifvmode
      \if@inlabel
        \global\setbox\@labels\hbox{%
          \llap{\a@lab\let\a@lab\relax
                \kern\@totalleftmargin\kern\marginparsep}%
          \box\@labels}%
      \fi
    \fi}
\fi
\documentclass[12pt,letterpaper]{article}
\usepackage{amsmath,amssymb,array,calc,rotating,epsfig,psfrag}
\usepackage[nosort]{cite}
\ifnum\draftcontrol=1
\fi
\renewcommand\baselinestretch{1.25}
\renewcommand\section{\@startsection {section}{1}{\z@}%
                                   {-3.5ex \@plus -1ex \@minus -.2ex}%
                                   {2.3ex \@plus.2ex}%
                                   {\normalfont\large\bfseries}}
\renewcommand\subsection{\@startsection{subsection}{2}{\z@}%
                                     {-3.25ex\@plus -1ex \@minus -.2ex}%
                                     {1.5ex \@plus .2ex}%
                                     {\normalfont\normalsize\bfseries}}
\renewcommand\subsubsection{\@startsection{subsubsection}{3}{\z@}%
                                     {-3.25ex\@plus -1ex \@minus -.2ex}%
                                     {1.5ex \@plus .2ex}%
                                     {\normalfont\normalsize\it}}

\numberwithin{equation}{section}

\def\revise#1       {\marginpar{\rule{2mm}{1cm} #1}}

\def\R{{\rm R}}

\def\sqr#1#2{{\vcenter{\vbox{\hrule height.#2pt  
 \hbox{\vrule width.#2pt height#1pt \kern#1pt
 \vrule width.#2pt}\hrule height.#2pt}}}}

\def\yboxit#1#2{\vbox{\hrule height #1 \hbox{\vrule width #1
\vbox{#2}\vrule width #1 }\hrule height #1 }}
\def\fillbox#1{\hbox to #1{\vbox to #1{\vfil}\hfil}}
\def\ybox{{\lower 1.3pt \yboxit{0.4pt}{\fillbox{8pt}}\hskip-0.2pt}}

\def\comments#1{}

\def\Tr{{{\rm Tr~ }}}

\def\CL{{\cal L}}
\def\CM{{\cal M}}
\def\CN{{\cal N}}
\def\CO{{\cal O}}

\def\II{\relax{I\kern-.10em I}}

\def\IZ{\relax\ifmmode\mathchoice
{\hbox{\cmss Z\kern-.4em Z}}{\hbox{\cmss Z\kern-.4em Z}}
{\lower.9pt\hbox{\cmsss Z\kern-.4em Z}}
{\lower1.2pt\hbox{\cmsss Z\kern-.4em Z}}\else{\cmss Z\kern-.4em
Z}\fi}
\def\IB{\relax{\rm I\kern-.18em B}}
\def\IC{{\relax\hbox{$\inbar\kern-.3em{\rm C}$}}}
\def\ID{\relax{\rm I\kern-.18em D}}
\def\IE{\relax{\rm I\kern-.18em E}}
\def\IF{\relax{\rm I\kern-.18em F}}
\def\IG{\relax\hbox{$\inbar\kern-.3em{\rm G}$}}
\def\IGa{\relax\hbox{${\rm I}\kern-.18em\Gamma$}}
\def\IH{\relax{\rm I\kern-.18em H}}
\def\II{\relax{\rm I\kern-.18em I}}
\def\IK{\relax{\rm I\kern-.18em K}}
\def\IP{\relax{\rm I\kern-.18em P}}
\def\inbar{\,\vrule height1.5ex width.4pt depth0pt}

\font\cmss=cmss10 \font\cmsss=cmss10 at 7pt
\def\IR{\relax{\rm I\kern-.18em R}}

\def\BZ{{\mathbb {Z}}}
\def\BP{{\mathbb {P}}}
\def\BC{{\mathbb {C}}}

\def\lp10{l_P^{10}}
\def\lp11{l_P^{11}}
\newcommand{\nc}{\newcommand}
\nc{\rnc}{\renewcommand}
\nc{\beq}{\begin{equation}}
\nc{\eeq}{\end{equation}}
\nc{\ntwo}{${\cal N}=2$}
\nc{\nOne}{${\cal N}=1$}
\nc{\hs}{\hspace{0.2in}}
\nc{\WP}{\mathbb{WP}}
\nc{\slag}{special Lagrangian}
\nc{\cn}{\C^n}
\nc{\rn}{\R^n}

\nc{\Wtree}{W_{\hbox{\small tree}}}
\nc{\Weff}{W_{\hbox{\small eff}}}
\nc{\Wkw}{{W_{\hbox{\tiny KW}}}}
\nc{\Wpw}{{W_{\hbox{\tiny PW}}}}
\nc{\lkw}{{\lambda_{\hbox{\tiny KW}}}}
\nc{\lpw}{{\lambda_{\hbox{\tiny PW}}}}
\def\Bid{{\mathchoice {\rm {1\mskip-4.5mu l}} {\rm
{1\mskip-4.5mu l}} {\rm {1\mskip-3.8mu l}} {\rm {1\mskip-4.3mu l}}}}

\def\bea{\begin{eqnarray}}
\def\eea{\end{eqnarray}}
\nc{\ra}{\rightarrow}
\nc{\non}{\nonumber}

\catcode`\@=12

\begin{document}

\title{\Large \bf 
$\CN=1$ Field Theories and Fluxes in
IIB String Theory}

\pubnum{%
ILL-(TH)-03-11 \\
USC-04-01 \\
hep-th/0401141}
\date{January 2004 \\
Email: {\tt rcorrado@uiuc.edu, halmagyi@usc.edu}}

\author{Richard Corrado${}^\flat$ and  
Nick Halmagyi${}^{\sharp}$ \\[0.4cm]
\it ${}^\flat$Department of Physics\\
\it University of Illinois at Urbana-Champaign \\
Urbana, IL 61801, USA \\[0.2cm]
\it ${}^\sharp$Department of Physics and Astronomy\\
\it University of Southern California \\
\it Los Angeles, CA 90089, USA 
}

\Abstract{Deformation of $\CN=2$ quiver gauge theories by adjoint
masses leads to
fixed manifolds of $\CN=1$ superconformal field theories. We elaborate
on the role of the complex three--form flux in the IIB duals to these 
fixed point
theories, primarily using field theory techniques. We study the moduli
space at a fixed point and find that 
it is either the two (complex) dimensional ALE space or three--dimensional 
generalized conifold, depending on the type of
three-form flux that is present. We describe the exactly marginal
operators that parameterize the fixed manifolds and find the operators
which preserve the dimension of the moduli space.  We also study
deformations by arbitrary superpotentials $W(\Phi_i)$ for the
adjoints. We invoke 
the $a$--theorem to show that there are no dangerously irrelevant
operators like $\Tr\Phi_i^{k+1}$, $k>2$ in the $\CN=2$ quiver gauge
theories. The moduli space of the IR fixed point theory generally
contains orbifold singularities if  $W(\Phi_i)$ does not give a mass
to the adjoints. Finally we examine some nonconformal $\CN=1$ quiver 
theories. We find evidence that the moduli space at the endpoint of a 
Seiberg duality cascade is always a three--dimensional generalized conifold. 
In general, the low--energy theory receives quantum corrections. In several 
non--cascading theories we find that the moduli space is a generalized conifold
realized as a monodromic fibration.
}

\enlargethispage{1.5cm}

\makepapertitle

\vfill \eject 

\tableofcontents

\body

%\version\versionno almost there

%%%%%%%%%%%%%%%%%%%%%%%%%%%%%%%%%%%%%%%%%

\baselineskip16pt
\section{Introduction}

One of the biggest successes of string theory in the last 10 years has
been the development of the correspondence between gauge theories and
gravity. One aspect of this correspondence is the geometric
engineering of gauge theories by considering string theory
backgrounds in certain limits. In this paper, we will discuss various
aspects of some of the  $\CN=1$ gauge theories in four dimensions that
can be obtained by deforming the theories on D-branes at orbifold
singularities in IIB string theory.

The AdS/CFT correspondence~\cite{Maldacena:1998re,Aharony:1999ti}
provides one concrete route toward the theories we study in this
paper. One begins with D3-branes at the ALE orbifold $\BC^2/\Gamma$,
$\Gamma\subset SU(2)$, whose near horizon limit  is given by IIB
supergravity in the AdS$_5\times S^5/\Gamma$
background~\cite{Kachru:1998ys}. The dual conformal field theory in
this case is an $\CN=2$ quiver gauge
theory~\cite{Douglas:1996sw}. Deformations of this theory by relevant
operators 
will drive the theory to new conformal fixed points. One class of
relevant operators are certain mass terms for the adjoint chiral
fields~\cite{Klebanov:1998hh,Gubser:1998ia}, which drive the theory to
fixed points with $\CN=1$ superconformal invariance. These fixed point
theories are parameterized by superpotential couplings $h_i$. Generally, these
deformations lead to manifolds of fixed points parameterized by the
ratios $h_i/h_j$, so the  manifold of fixed points is a complex
projective space $\BP^{n-1}$~\cite{Corrado:2002wx}. 

The geometry dual to these fixed points is that of
AdS$_5\times X_5$, where $X_5$ is the 
base of a generalized
conifold~\cite{Klebanov:1998hh,Gubser:1998ia,Morrison:1998cs,%
Cachazo:2001gh,Cachazo:2001sg}.  By a generalized conifold, we mean the 
three complex--dimensional manifold obtained by fibering an ALE space 
over the complex line~\cite{KatzMorrison}.  In the field theory, this 
fibration arises directly from adding the mass terms. The orbifold 
singularities of the ALE 
space are replaced by conifold singularities at points on the line. The 
masses themselves correspond to complex structure moduli of the 
generalized conifold.

The geometry is, however, not the complete story, as a  mass
deformation with $\sum_i h_i^{-1}\neq 0$ also introduces flux for the
complex IIB 3--form field strength. In the presence of this flux, the
metric on the generalized conifold will differ from the Ricci--flat
one. The relation of these fluxes to geometric deformations was
described in a framework of 5D gauged supergravity
in~\cite{Corrado:2002wx}. There the map between mass terms and fields
in the untwisted and twisted sectors of the orbifold string theory was
examined. The symmetries of the manifold of fixed points in the dual
gauge theory were related to the duality symmetries of the orbifold
theory.

Our motivation for the present work was to shed further light on this
larger picture of 3--form flux in these theories. In
Section~\ref{sec:fluxes}, we review the results
of~\cite{Corrado:2002wx} on the correspondence between untwisted and
twisted sector fields and the relevant operators which drive the RG
flows to the manifold of fixed points.  Some solutions are known for the
supergravity duals to the fixed points generated by purely twisted
deformations. For $A_1$, the gravity dual for a purely twisted
deformation is just IIB on the conifold~\cite{Klebanov:1998hh}. More 
generally, for the $A_{n-1}$ case, there is some information for the 
duals to purely twisted deformations~\cite{Gubser:1998ia}. In the case 
of a purely untwisted deformation, the solution is an orbifold of the 
squashed sphere with 3--form flux solution 
of~\cite{Pilch:2000ej,Pilch:2000fu}.  However very little detail is
available when both twisted and untwisted deformations are
present. Hopefully, our field theory results can prove useful in
shedding light on the structure of the gravity duals of these fixed
points.  

In Section~\ref{sec:massdef},
we give a purely field theoretic analysis of the fixed points. One new
tool we employ is the $a$ maximization technique
of~\cite{Intriligator:2003jj} to compute the exact $\CN=1$ 
superconformal $U(1)_R$ R-charges.   
We then compute the moduli space of the scalar fields in the gauge
theory at the $\CN=1$ 
fixed points. The primary result is that, whenever the mass
deformation includes a deformation corresponding to the untwisted
sector, the moduli space is just two complex--dimensional. Namely, it
is the ALE orbifold, possibly resolved by D--terms. The existence of a
fixed point requires that F--terms in the quiver theory are zero. When
the mass deformation is purely in the twisted sector, corresponding to
the condition that $\sum_i h_i^{-1} =0$, we recover the
generalized conifold as the moduli space, in agreement with earlier
works~\cite{Gubser:1998ia,Cachazo:2001gh}.  We refer to the surface
$\sum_i h_i^{-1} =0$ on the manifold of fixed points as the ``conifold
subspace.'' This submanifold has the geometry of a
$\BP^{n-2}\subset \BP^{n-1}$.

The fact that untwisted mass deformations reduce the dimension of the
moduli space of scalars is intimately related to the presence of
3--form flux in 
the string dual. It is known that this 3--form flux can generate a
non--zero potential for a probe D3--brane in this
geometry~\cite{Johnson:2000ic}. The moduli space of the gauge theory
is located at the minimum of this potential. This corresponds to a
point on the complex line used to realize the generalized conifold as
an ALE fibration, so the moduli space reduces to just the ALE fiber. 
The gauge theory results suggest that this ALE space has a
resolution, but that the untwisted 3--form flux presents an
obstruction to complex structure deformations.

Another main result of the paper is Section~\ref{sec:marginal}, where
we discuss the exactly marginal operators that parameterize the
manifolds of fixed points.  We find that a general exactly marginal
perturbation will take a point on the conifold subspace off to a point
where the dual string background has nonzero 3--form flux. From the
moduli space perspective, the generic perturbation lifts one flat
(complex) direction, reducing the dimension of moduli space.  We find
the form of the operators which preserve the conifold subspace, which
depend on the initial position on the fixed manifold.

In Section~\ref{sec:gendef}, we analyze deformations of the quiver
gauge theories by arbitrary polynomial superpotentials for the adjoint
chiral fields, $W(\Phi_i)$,
following~\cite{Cachazo:2001gh,Cachazo:2001sg}. These deformations
include operators which are irrelevant at the $\CN=2$ UV fixed
point. Our main result is that deformations by irrelevant operators do
not lead to new conformal fixed points. We show this by demonstrating
that these fixed points, if they existed, would lead to violations of
the conjectured
$a$--theorem~\cite{Anselmi:1998am,Anselmi:1998ys}. Specifically, using
a result 
of~\cite{Cachazo:2001sg} on the central charge of the candidate fixed
points for monomial deformations $\Phi_i^{k+1}$, we show that the
flows away from these point generated by relevant perturbations
$\Phi_i^{k'+1}$, $k'<k$, would violate the $a$--theorem. We then
compute the moduli spaces for these theories by assuming that in the
IR the theory is sitting at a critical point of $W(\Phi_i)$. Then we
can use the effective mass term for perturbations around the critical
point to recover the moduli spaces with data specified by the
effective masses.

Nonconformal field theories can also be studied by adding fractional
branes to the string backgrounds describing the conformal field
theory~\cite{Gubser:1998fp,Klebanov:1999rd,Klebanov:2000nc,%%
Klebanov:2000hb,Cachazo:2001gh,Cachazo:2001sg}.
In these nonconformal field theories, the superpotential generally
receives corrections. Correspondingly, the moduli space is a deformed
version of the moduli space of the related conformal field theory. In
Section~\ref{sec:nonconformal}, we generalize some of the field theory
discussion
of~\cite{Klebanov:2000hb} to theories which arise from untwisted
deformations of $\CN=2$ theories. We discuss the RG flow and Seiberg
duality. 

These theories are characterized by couplings $h_i$ in a quartic
superpotential. In analogy with the conformal theories, when 
$\sum_i h_i^{-1}\neq 0$, the moduli space of scalars is a two
complex--dimensional ALE space, whereas when $\sum_i h_i^{-1}=0$, it
is a generalized conifold.  We find an indication that, in theories
which undergo a duality cascade, the cascade maps the theory onto a
new theory which is in the conifold subspace. When $\sum_i
h_i^{-1}\neq 0$, the moduli space is growing an extra dimension at the
end of the cascade. However, quantum corrections to the superpotential
could change this result. 

We also study some examples in which the corrections to the
superpotential are known. These theories are too simple to undergo a
duality cascade. The quantum corrections lead to a complex deformation
of the ALE or generalized conifold moduli space. In particular, the
deformation of 
the generalized conifold leads to a monodromic fibration structure. 
 
Section~\ref{sec:quivers} contains an introduction to $\CN=2$ quiver
gauge theories, in order to set up notation and some of the
computations made in later sections. We conclude with a discussion of
our results and interesting future directions of research.

\section{Review of $\CN=2$ Quiver Gauge Theories}
\label{sec:quivers}

The $\CN=2$ quiver gauge theories can be easily described using the
language of $\CN=1$ superfields. The gauge group $G$ is a product
$G=\times_{i=1}^n G_i$ of Lie groups $G_i$. The $\CN=2$
vector multiplets contribute matter consisting of $\CN=1$ chiral
fields $\Phi_i$ in the adjoint of $G_i$. For each factor of the gauge
group, we add a vertex to the quiver diagram. Additional matter can come in
the form of $\CN=2$ hypermultiplets, which decompose into pairs of
$\CN=1$ chiral fields $A_{ij}$ and $B_{ji}$. We will consider the case that
the hypermultiplets lie in bifundamental representations of the gauge
group, {\it i.\ e.},  $A_{ij}$ is in the 
$(\mathbf{N}_i, \bar{\mathbf{N}}_j)$ and $B_{ji}$ is in the 
$(\bar{\mathbf{N}}_i, \mathbf{N}_j)$. For each field $A_{ij}$ we draw
an oriented line from the vertex $i$ to the vertex $j$, while for each
$B_{ji}$ we draw an oriented line from $j$ to $i$. It is possible to
have several ``flavors'' of fields connecting two vertices with the
same orientation. To avoid cluttering notation, we will resist adding
flavor indices to our discussion. 

There is a unique renormalizable
superpotential allowed in these theories, 
\begin{equation}
\Wtree =  \sum_{i\neq j} \lambda_i\,\text{Tr}_{\mathbf{N}_i}\, a_{ij} \Phi_i 
\left(A_{ij} B_{ji}-B_{ij}A_{ji}\right),
\label{eq:Neqtwotree}
\end{equation}
where $a_{ij} \in \{0,1\}$ are the elements of the (symmetric)
adjacency matrix of 
the quiver. The theory with this superpotential admits  $\CN=2$ SUSY
when the 
couplings $\lambda_i=q_ig_i$, where $g_i$ is the gauge coupling of $G_i$
and $q_i$ is the charge of $A_{ij}$ under $G_i$. There is an
$SU(2)\times U(1)_R$ R--symmetry under which $\Phi_i$ are in the
$\mathbf{1}_2$ and the pairs $(A_{ij},B_{ji})$
correspond to a $\mathbf{2}_0$. However, this $SU(2)$ is generally not
manifest in a superpotential written in terms of $\CN=1$ superfields,
such as~\eqref{eq:Neqtwotree}.  

When the
$G_i$ are not simple, F and D--terms can be added to the Lagrangian as
well,
\begin{equation}
\CL_{F,D} =  \sum_{i} \left[ \int d^4\theta\, d_i\, \Tr V_i + 
\int d^2\theta 
  f_i\, \Tr \Phi_i + \text{c.c.}\right],
\label{eq:NeqtwoFandD}
\end{equation}
where $d_i$ and $f_i$ are (complex and real, respectively)
parameters. 
The supersymmetric vacua of these theories are the solutions to the D and
F--flatness conditions, which read
\begin{equation}
\begin{split}
& \sum_j a_{ij}q_i  
\left( A_{ij} A_{ij}^\dagger - B_{ij}^\dagger B_{ij}
-  A_{ji}^\dagger A_{ji} + B_{ij} B_{ij}^\dagger\right)
 +d_i =0, \\ 
& \sum_j a_{ij} \left(A_{ij} B_{ji}-B_{ij}A_{ji}\right) +f_i =0,\\
& \sum_j \left(A_{ij} \Phi_j - \Phi_i A_{ij}\right) =0, \\
& \sum_j \left(B_{ij}\Phi_j -\Phi_i B_{ij}\right) =0. 
\end{split}
\label{eq:NeqtwoFandDflat}
\end{equation}
Consistency of the first equations with the adjacency of the quiver
nodes will require that $\sum_i f_i =\sum_i d_i =0$.

The one--loop exact beta--function for the gauge coupling $g_i$ is 
\begin{equation}
\beta(g_i) = - \frac{g_i^3}{16\pi^2}\, 
\frac{\left(2+\tfrac{1}{2}\gamma_{\Phi_i}\right)T(G_i) 
- \tfrac{1}{2}\sum_j a_{ij} N_j 
\left(2-\tfrac{1}{2}\gamma_{A_{ij}}- \tfrac{1}{2}\gamma_{B_{ji}}\right)}{
1-\frac{g_i^2T(G_i)}{8\pi^2}}.
\label{eq:genbeta}
\end{equation}

Of particular interest to us are the class of $\CN=2$ quiver gauge
theories that can be obtained by studying D-branes at orbifold
singularities~\cite{Douglas:1996sw}. By placing $N$ D3-branes
transverse to the orbifold 
$\BC^2/\Gamma$, with $\Gamma$ a finite subgroup of $SU(2)$, a 4D
$\CN=2$ gauge theory is obtained on the worldvolume. These $\Gamma$
have an $A$--$D$--$E$ classification. The $U(N)$ gauge theory is
broken to a product $\prod_i U(N_i)$, where $N=\sum_i N_i$ and the
$N_i$ are in 1--1 correspondence with the vertices of the affine (or
extended) Dynkin diagram. These vertices are the simple roots
$\alpha_i$, $i=1,\ldots,n-1$ plus the extended root
$\alpha_0=-\sum_{i=1}^{n-1} \alpha_i$ of the simply--laced
$A$--$D$--$E$ algebra. The inner product on the 
roots determines the adjacency matrix of the quiver diagram as 
$a_{ij}=2\delta_{ij}-\widehat{C}_{ij}$
where $\widehat{C}_{ij}$ is the extended Cartan matrix. 
The hypermultiplets have charge
$q_i=1$ and their moduli space describes the resolved ALE space
$\widetilde{\BC^2/\Gamma}$, while the scalars $\Phi_i$ describe the
remaining $\BC$ transverse directions. 

The leading beta--function coefficient for $g_i$ is 
\begin{equation}
b_0^{(i)} = 2 T(G_i) 
-  \sum_j a_{ij} N_j = \sum_j \widehat{C}_{ij} N_j. \label{eq:betazero}
\end{equation}
When only
regular D3-branes are present, $N_i = k_i \widetilde{N}$, where the $k_i$
are the Dynkin labels of the algebra. Since 
$\sum_j \widehat{C}_{ij} k_j =0$ as a Lie algebra identity, this field
theory can be superconformal.  
If 
there are $r_i$ fractional branes wrapping the $i^{th}$ homology cycle
of the ALE,  
then $N_i = k_i \widetilde{N} +r_i$. Now 
$b_0^{(i)}= \sum_j \widehat{C}_{ij} r_j$ and the free field theory is not
conformal. If $b_0^{(i)}>0$, the (simple part of the) gauge group
$G_i$ is asymptotically free.

Let us examine the quiver gauge theory on regular D3--branes in some
more detail. We set $N_i = k_i \widetilde{N}$. We apply the conditions for
conformal invariance 
following~\cite{Leigh:1995ep}. Similar discussions for these theories
appear in~\cite{Klebanov:1998hh,Gubser:1998ia,Cachazo:2001sg}. Then 
vanishing of the exact beta functions for the gauge
couplings~\eqref{eq:genbeta} lead to the conditions 
\begin{equation}
k_i \gamma_{\Phi_i}(\tau_i,\lambda_i) 
+ \tfrac{1}{2}\sum_j a_{ij} k_j 
\left(\gamma_{A_{ij}}(\tau_i,\lambda_i)
+\gamma_{B_{ji}}(\tau_i,\lambda_i)\right) =0, \label{eq:betags}
\end{equation}
while vanishing of the beta--functions for the $\lambda_i$ require
that 
\begin{equation}
\begin{split}
\gamma_{\Phi_i}(\tau_i,\lambda_i) 
+ \sum_j a_{ij}  
\left(\gamma_{A_{ij}}(\tau_i,\lambda_i)
+\gamma_{B_{ji}}(\tau_i,\lambda_i)\right) =0, \\
\gamma_{\Phi_i}(\tau_i,\lambda_i) 
+ \sum_j a_{ij}  
\left(\gamma_{A_{ji}}(\tau_i,\lambda_i)
+\gamma_{B_{ij}}(\tau_i,\lambda_i)\right) =0,
\end{split} \label{eq:betals}
\end{equation}
where the $\tau_i = \vartheta_i + 4\pi i/g_i^2$ are the complexified
gauge couplings. 
The R--symmetry component $SU(2)$ that acts on the hypermultiplets
requires that $\gamma_{B_{ji}} = \gamma_{A_{ij}}$, while compatibility
of~\eqref{eq:betags} with the first of~\eqref{eq:betals} requires that 
$\gamma_{A_{ij}}\equiv \gamma_A$ are the same for all
$i,j$. Furthermore, the two equations in~\eqref{eq:betals} can be used
to show that $\gamma_{\Phi_i}=\gamma_\Phi$ for all $i$.  Then
conformal invariance requires that 
\begin{equation}
\gamma_\Phi(\tau_i,\lambda_i) + 2 \gamma_A(\tau_i,\lambda_i) =0.
\label{eq:anomconst}
\end{equation}
This describes a fixed surface $\lambda_i = \lambda_i(\tau_i)$, which
we will denote $\CM_\tau^{(n)}$. Its structure is discussed
in~\cite{Witten:1997sc,Dorey:2001qj}. 

We will now determine the anomalous dimensions by using the
$a$--maximization technique 
of~\cite{Intriligator:2003jj} to compute the exact $\CN=1$
superconformal $U(1)_R$ R-charges. Applications of this method to SQCD
with adjoint matter were discussed
in~\cite{Kutasov:2003iy,Intriligator:2003mi}, which contain important
refinements. We use the relation between the anomalous dimension and
R-charge, $\gamma=\tfrac{3}{2}R-1$, to write
\begin{equation}
R_\Phi + 2 R_A -2 =0.
\end{equation}
Next we compute the $a$--charge, which is given
by~\cite{Anselmi:1998am,Anselmi:1998ys}
\begin{equation}
a = \frac{3}{32} \left( 3 \Tr R^3 - \Tr R\right),
\end{equation}
where the traces are performed over all fermions in the theory. For
the quiver theories 
\begin{equation}
\begin{split}
a = & \frac{3}{32} \left[ 2\sum_i N_i^2 
+ \sum_i N_i^2  \left( 3 (R_{\Phi_i}-1)^3 - (R_{\Phi_i}-1)
\right) \right.
\\
&  \hspace{1cm} +   \sum_{i<j} a_{ij} N_i N_j 
\Bigl( 3 (R_{A_{ij}}-1)^3 - (R_{A_{ij}}-1)  \\
& \left.\phantom{ \sum_{i<j}} \hspace{3cm}
+ 3 (R_{B_{ji}}-1)^3 - (R_{B_{ji}}-1)\Bigr) 
\right].
\end{split} \label{eq:acharge}
\end{equation}
Maximizing this with respect to the constrained R-charges leads to
\begin{equation}
R_{\Phi_i} =  R_{A_{ij}} = R_{B_{ji}}  =\frac{2}{3}.
\label{eq:NtwoRcharges}
\end{equation}
So $a$--maximization actually requires that the anomalous dimensions
vanish, which is much stronger than the condition of conformal
invariance~\eqref{eq:anomconst}.  This occurs on the  $\CN=2$
fixed line $\lambda_i=g_i$.  
The result~\eqref{eq:NtwoRcharges} agrees with that obtained by
studying each gauge factor 
independently as a theory with $N_c=N$, $N_f=2N$ and
following~\cite{Seiberg:1995pq}. 

\section{Flux Deformation of IIB Orbifolds}
\label{sec:fluxes}

The near horizon limit of the theory on $N$ D3-branes at the orbifold
$\BC^2/\Gamma$  is given by IIB
supergravity in the AdS$_5\times S^5/\Gamma$
background~\cite{Kachru:1998ys}. Here since 
$\Gamma\subset SU(2)$, there is a singularity of $S^5/\Gamma$
corresponding to a fixed circle. The isometry of $S^5/\Gamma$ is
$SU(2)\times U(1)$, which is the R-symmetry of the dual quiver gauge
theory. 
The spectrum of this theory is discussed
in~\cite{Oz:1998hr,Gukov:1998kk}. Among the states which are massless
on AdS$_5$ are $n=\text{ord}(\Gamma)$ 5D tensor multiplets. 

One tensor multiplet comes
from the untwisted sector of the orbifold theory and is dual to
the chiral primary operator 
\begin{equation}
\CO = \Tr \sum_i\,  \phi_i^2 \label{eq:untwistop}
\end{equation}
and its descendants, where $\phi_i$ is the scalar component of the
superfield $\Phi_i$.  There are 5 scalars in this tensor
multiplet~\cite{Gunaydin:1986cu}  and
they have $SU(2)\times U(1)$ charges $\mathbf{1}_4$, $\mathbf{3}_2$,
and $\mathbf{1}_0$.  There is also a conjugate tensor multiplet
with scalars $(\mathbf{1}_{-4},\mathbf{3}_{-2},\mathbf{1}_0)$
corresponding to the antichiral operator, 
$\Tr \sum_i\, (\phi_i^\dagger)^2$, conjugate
to~\eqref{eq:untwistop}.  These scalars are built from many types of 10D
fields. The $\mathbf{1}_{\pm 4}$ states are linear combinations of the
lowest harmonics of metric (${h^\alpha}_\alpha$) and 4-form potential
($a_{\alpha\beta\gamma\delta}$) degrees of freedom~\cite{Kim:1985ez}.
The $\mathbf{3}_{\pm 2}$ arise from the lowest harmonic of the complex
2-form potential components ($a_{\alpha\beta}$), while the
pair $2(\mathbf{1}_0)$ is the complex axion--dilaton.

The remaining $n-1$ tensor multiplets come from
the twisted sectors and are dual to the differences 
\begin{equation}
\CO_i = \Tr \Bigl( \phi_i^2 - \phi_{i-1}^2 \Bigr)
\label{eq:twistedops}
\end{equation}
and descendants.
Including the duals to the antichiral operators, the
$\mathbf{1}_{\pm 4}$, $2(\mathbf{1}_0)$ scalars 
are linear combinations of harmonics of the periods of the complex
2-form potential over the compact 2-cycles of the $\BC^2/\Gamma$
orbifold. The $\mathbf{3}_{\pm 2}$ states are the lowest harmonics of
the moduli associated with varying the sizes of the compact 2-cycles,
{\it i.e.}, they are the blow-up modes.

The operators~\eqref{eq:untwistop}, \eqref{eq:twistedops} are
relevant, as are their level two descendants, which are built from
untwisted and 
twisted sums of fermion bilinears $\Tr\chi_i\chi_i$. Deformation of the  
conformal field theory by them will generate flows to an $\CN=1$
conformal field
theory~\cite{Klebanov:1998hh,Gubser:1998ia,Bergman:2001qi,%
Dorey:2001qj,Cachazo:2001sg,Corrado:2002wx}.
We will discuss the field theoretic properties of these fixed points
in the next section. For now, we would like to discuss aspects of the
dual geometry,
following~\cite{Klebanov:1998hh,Gubser:1998ia,Cachazo:2001sg,Corrado:2002wx}.  

The description of the RG flows generated by~\eqref{eq:untwistop},
\eqref{eq:twistedops} (and descendants) within the effective 5D 
$\CN=4$ $SU(2)\times
U(1)$ gauged supergravity was described
in~\cite{Corrado:2002wx}. There it was found that the 5D dynamics is
symmetric under an $SU(n)$ acting on the tensor multiplets. In
particular, this can be used to map a flow generated by a generic
initial condition to one involving only the untwisted sector
scalars. 

The flows generated in the untwisted sector are completely
analogous to those studied in the theory on $S^5$, without the
orbifold, by~\cite{Freedman:1999gp}.  In particular, the flows
generated by the fermion bilinears, corresponding to the
$SU(2)$--singlet component of the $\mathbf{3}_{2}$, can be precisely
mapped to the $SU(2)\times U(1)$ preserving flow
of~\cite{Freedman:1999gp}.  The
5D flow solution of~\cite{Freedman:1999gp} was lifted to 10D
in~\cite{Pilch:2000ej,Pilch:2000fu}. The 10D solution involves a
stretched and squashed metric on $S^5$, together with background
fluxes for the complex 3--form and 5--form field strengths. 
In~\cite{Corrado:2002wx} it was argued that, since the 3 and 5--forms
are invariant under $\BZ_n$, the orbifold of the solution
in~\cite{Pilch:2000ej,Pilch:2000fu} is the 10D lift of the
flow generated purely by the untwisted operator~\eqref{eq:untwistop}
in the corresponding quiver gauge theory.

The 10D geometries of the flows generated by purely twisted sector
operators~\eqref{eq:twistedops} are very different from that generated
by the untwisted sector, however.  Since the 
$\mathbf{3}_{2}$ are blow-up modes, these flows correspond to 
desingularizing the orbifold
singularity~\cite{Klebanov:1998hh,Gubser:1998ia}. Strictly speaking,
the orbifold singularity is deformed. Introducing complex coordinates
$x$, $y$, $z$, and $t$, if the ALE space is the $A_{n-1}$ curve
\begin{equation}
xy = z^n
\end{equation}
in $\BC^3$, the effect of an single twisted sector operator  is to
deform this to
\begin{equation}
xy = z^{n-2} (z-\zeta \, t)(z+\zeta \, t),
\end{equation}
which is an example of a generalized conifold.
Part of the orbifold singularity at $x=y=z=0$ has been replaced with
a  conifold singularity at $x=y=z=t=0$. The 2--sphere corresponding to
the twisted sector of the operator now has area $|\zeta\, t|$. The ALE
space is said to be fibered over the line parameterized by $t$. A
generic twisted deformation leads to the curve
\begin{equation}
xy = \prod_i^n (z-\zeta_i \, t) , ~~ \sum_i \zeta_i =0.
\label{eq:genconfirst}
\end{equation}
At the fixed 
point, the near horizon solutions involve only 5--form flux.
Nevertheless, the 5D symmetry suggests that there should be some map
between IIB 
fields corresponding to the different endpoints of the RG
flows. However, metrics for the endpoints of these flows are only
known in a small 
neighborhood of the conifold singularities~\cite{Gubser:1998ia}. 

When some untwisted sector flux is added, a 3--form flux is generated
on the generalized conifold~\eqref{eq:genconfirst}.  This leads to a
potential on the worldvolume of a probe brane in the generalized
conifold geometry. We will find that the moduli space of the gauge
theory on the probe brane is just the ALE fiber. This space
corresponds to the minimum of the potential on the probe. 

\subsection{An illustration for $A_1$}
\label{sec:illustrate}

The above picture is easiest to illustrate in the case of the
$\widehat{A}_1$ quiver theory. This theory has two adjoint scalars
$\Phi_1, \, \Phi_2$ and there are two possible mass deformations.  The
untwisted deformation~\eqref{eq:untwistop} corresponds to adding the term
\begin{equation}
W_u = \Phi_1^2 + \Phi_2^2 \label{eq:aoneuntwist}
\end{equation}
to the $\CN=2$ superpotential. The geometry dual to the endpoint of
the flow generated by~\eqref{eq:aoneuntwist} is a $\BZ_2$ orbifold of
the solution of~\cite{Pilch:2000ej}. As remarked above, the metric is
that of AdS$_5$ times a stretched and squashed $S^5/\BZ_2$ and there is
nonzero complex 3--form and 5--form flux. There is a $\BZ_2$ orbifold
singularity which is a fixed line on the $S^5/\BZ_2$. We will call
this fixed point the PW point. 

The twisted deformation~\eqref{eq:twistedops} is the term
\begin{equation}
W_t = \Phi_1^2 - \Phi_2^2. \label{eq:aonetwist}
\end{equation}
The gravity dual to the fixed point is now AdS$_5$ times the base
manifold $T^{11}$ of the 
conifold~\cite{Klebanov:1998hh}. There is 5--form flux, but no 3--form
flux. The orbifold singularity on $S^5$ is removed and $T^{11}$ is
smooth. We call the corresponding fixed point the KW point.

A general deformation
\begin{equation}
W = \frac{m_1}{2} \Phi_1^2 + \frac{m_2}{2} \Phi_2^2 
\label{eq:aonegeneral}
\end{equation}
actually describes a point on the complex projective plane $\BP^1$,
which can also be identified with a 2--sphere. The reason is that the
overall mass scale decouples in the IR, so the ratio $m_2/m_1$
specifies the fixed point. The gravity dual for a general
deformation~\eqref{eq:aonegeneral} has not been constructed, but it
involves both adding 3--form flux and desingularizing the orbifold
singularity. 

\begin{figure}[ht] 
\begin{center}      
\epsfxsize=9cm \epsfbox{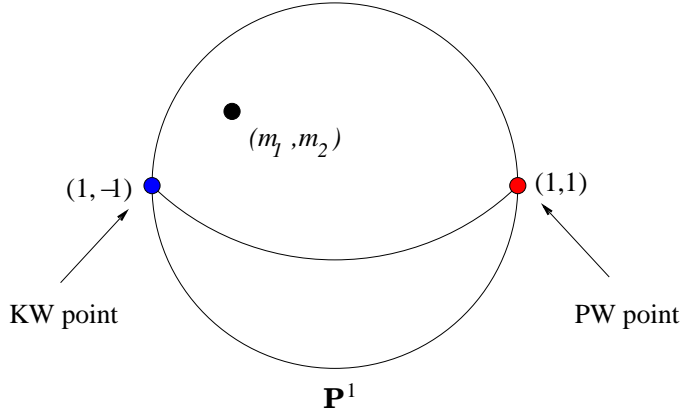}    
\end{center}
\caption{The $\BP^1$ of fixed points for the adjoint mass
deformation~\eqref{eq:aonegeneral} of the $A_1$ quiver theory. The PW
and KW points are indicated, as well as the generic point
$(m_1,m_2)$.}   
\label{fig:Aone}  
\end{figure}

In terms of the homogeneous coordinates
$(m_1,m_2)$, the PW point is $(1,1)$, while $(1,-1)$ is the KW
point. In Figure~\ref{fig:Aone}, we represent the $\BP^1$ of fixed
points with a 2--sphere and indicate the PW and KW points. 
We leave further discussion of the manifolds of fixed points to
Section~\ref{sec:marginal}. 

\section{The Mass Deformation and $\CN=1$ Fixed Points}
\label{sec:massdef}

For simplicity, let us consider the superconformal quiver gauge theory
on $nN$ regular 
D3--branes at the $\BC^2/\BZ_n$ singularity. This is the
$\widehat{A}_{n-1}$ theory. The gauge group is $G=U(N)^n$ and the
quiver corresponds to the affine $\widehat{A}_{n-1}$ Dynkin diagram. 
The adjacency matrix
is $a_{ij} = \delta_{i-1,j} + \delta_{i+1,j}$. We will use a slightly
simplified notation for the hypermultiplets and define 
$A_i \equiv A_{i,i+1}$, $B_i \equiv B_{i+1,i}$.  The superpotential is
\begin{equation}
\Wtree =  \sum_i \lambda_i\,\text{Tr}\,  \Phi_i 
\left(A_{i} B_{i}-B_{i-1}A_{i-1}\right).
\label{eq:WA}
\end{equation}

As discussed in Section~\ref{sec:fluxes}, deformations of this theory
by relevant operators generate flows to other conformal fixed
points. We will study the $\CN=1$ superconformal field
theories obtained by a deformation by mass terms for the adjoint
chiral fields 
\begin{equation}
W_\Phi = \Tr \sum_i \frac{ m_i}{2} \, \Phi_i^2. \label{eq:massdef}
\end{equation}
A purely twisted sector deformation will satisfy $\sum_i m_i=0$, while
$\sum_i m_i \neq 0$ indicates that the untwisted sector deformation is
present. 
The action on the hypermultiplet fields $(A_i,B_i)$ of the $SU(2)$
component of the $\CN=2$ R--symmetry is preserved 
by~\eqref{eq:massdef} and becomes a global symmetry of the $\CN=1$ theory.

For each field $\Phi_i$ of mass $m_i\neq 0$, as we flow to scales
$\mu< m_i$, we should integrate  
$\Phi_i$ out of the dynamics. The remaining theory is that of the
hypermultiplet fields and any massless $\Phi_i$, with the effective
superpotential
\begin{equation}
 W = \Tr \left[ -
\sum_{i|m_i\neq 0} \frac{h_i}{2} \, 
\left(A_{i} B_{i}-B_{i-1}A_{i-1}\right)^2
+ \sum_{i|m_i= 0}  \lambda_i\, \Phi_i 
\left(A_{i} B_{i}-B_{i-1}A_{i-1}\right)
\right],  \label{eq:Asuppot}
\end{equation}
where
\begin{equation}
h_i = \frac{\lambda_i^2}{m_i}. \label{eq:hdef}
\end{equation}
When $m_i\neq 0$, we will use~\eqref{eq:hdef} to eliminate $m_i$.

Note that the presence of nonzero F--terms in~\eqref{eq:WA} will
result in the addition of a term 
\begin{equation}
-  \Tr\sum_{i|m_i\neq 0} \frac{h_if_i}{\lambda_i} \, 
\left(A_{i} B_{i}-B_{i-1}A_{i-1}\right)
\label{eq:abmass}
\end{equation}
to~\eqref{eq:Asuppot}. This is a mass term and, were it to be nonzero,
the corresponding $(A_i,B_i)$ should be integrated out of the
low--energy theory. Generally it is expected that this theory will
continue to RG flow 
into the IR and there will be no fixed point. If all of the $m_i\neq
0$, then for the mass 
term to be absent the $f_i$ must satisfy
\begin{equation}
f_i = \frac{\lambda_i}{\lambda_{i-1}}\,\frac{h_{i-1}}{h_i}\, f_{i-1}. 
\end{equation}
For generic $\lambda_i$ and $m_i$, this condition will fail to hold,
so  we conclude that the $f_i=0$.
However, we note that
there is no such prohibition against adding D--terms to the theory.

Conformal invariance of~\eqref{eq:Asuppot} requires
that
\begin{equation}
\begin{split}
\left. \begin{matrix}
\hspace*{7mm}\gamma_{\Phi_i}(\tau_i,\lambda_i,h_i)  
+ \gamma_{A_i}(\tau_i,\lambda_i,h_i)  
+ \gamma_{B_i} (\tau_i,\lambda_i,h_i) =0 \\
\gamma_{\Phi_i}(\tau_i,\lambda_i,h_i)+\gamma_{A_{i-1}}(\tau_i,\lambda_i,h_i)  
+ \gamma_{B_{i-1}}(\tau_i,\lambda_i,h_i)  =0
\end{matrix} \right\}~&~\forall~ i|h_i=0,  \\
\left. \begin{matrix}
\hspace*{7mm} \gamma_{A_i}(\tau_i,\lambda_i,h_i) 
+ \gamma_{B_i}(\tau_i,\lambda_i,h_i)+ \frac{1}{2} =0 \\
\gamma_{A_{i-1}}(\tau_i,\lambda_i,h_i)  
+ \gamma_{B_{i-1}}(\tau_i,\lambda_i,h_i) +\frac{1}{2} =0
\end{matrix} \right\} ,~&~\forall~ i|h_i\neq 0.
\end{split} \label{eq:Anfixedpts}
\end{equation}
One caveat about~\eqref{eq:Anfixedpts} is that whenever 
\begin{equation}
h_i + h_{i-1} =0
\label{eq:oppcoup}
\end{equation}
for some $i$, then conformal invariance will require the weaker
condition that
\begin{equation}
\gamma_{A_i} + \gamma_{B_i} +\gamma_{A_{i-1}} + \gamma_{B_{i-1}}+1
=0. \label{eq:Anweakcond}
\end{equation}

The $SU(2)$ global symmetry requires that
$\gamma_{A_i}=\gamma_{B_i}\equiv \gamma_i$. Let us first consider the
case that all $m_i\neq 0$ and that~\eqref{eq:oppcoup} does not
occur. Then the superpotential~\eqref{eq:Asuppot} depends only on
$h_i$. The~\eqref{eq:Anfixedpts} require that
\begin{equation}
\gamma_i (\tau_i,h_i) + \frac{1}{4} = 0.
\label{eq:Anfixedptssymm}
\end{equation}
This is $n$ equations in $2n$ unknowns. There is a manifold of fixed
points~\cite{Gubser:1998ia,Corrado:2002wx}. Whenever at least one
$m_i$ is finite,  the
$h_i(\tau_i)$ form a 
$\BP^{n-1}$ manifold of fixed points\footnote{The geometry is
$\BP^{n-1}$ for both the bare $h_i$ and the renormalized couplings
$h_i(\tau_i)$, since~\eqref{eq:Anfixedptssymm} ensures a common
wavefunction renormalization for all of the quartic operators
in~\eqref{eq:Asuppot}. Therefore the ratios $h_i/h_j$ are not
renormalized.}, with inhomogeneous coordinates 
$h_i/h_j$. 

A special point is obtained when all $m_i\rightarrow\infty$. In this
limit the 
superpotential~\eqref{eq:Asuppot} vanishes. Each gauge theory is a
copy of SQCD with $N_f = 2N_c$ flavors, so it is in the conformal
window of~\cite{Seiberg:1994bz}. We therefore expect that this theory
also flows to a superconformal fixed point.

In the case that~\eqref{eq:oppcoup} holds, then some of the
equations~\eqref{eq:Anfixedptssymm} will be replaced by
\begin{equation}
\gamma_i (\tau_i,h_i)
+\gamma_{i-1} (\tau_i,h_i) + \frac{1}{2} = 0.
\label{eq:Anfixedptssymmtwo}
\end{equation}
We still have $n$ equations in $2n$ unknowns and the same picture of a
$\BP^{n-1}$ manifold of fixed points emerges.

When some $m_i=0$, we can apply cyclicity of the quiver
to~\eqref{eq:Anfixedpts} to find that 
\begin{equation}
\gamma_{\Phi_i}(\tau_i,\lambda_i,h_i) = \frac{1}{2}, ~~ m_i =0.
\end{equation}
These equations can be used to find the $\lambda_i(\tau_i,h_i)$, which
describe a manifold of the form $\CM_\tau^{(\nu)}$, where $\nu$ counts
the number of the $m_i=0$. Then
the equations~\eqref{eq:Anfixedptssymm}
or~\eqref{eq:Anfixedptssymmtwo} will determine the
$h_i(\tau_i)$, which now parameterize a $\BP^{n-\nu-1}$. 
This $\BP^{n-\nu-1} \subset \BP^{n-1}$ as the vanishing loci of sets
of $\nu$ of the $h_i^{-1}$ in the theory with 
$m_i \neq 0$~\cite{Corrado:2002wx}.

\subsection{Analysis by $a$--maximization}

We would like to analyze the fixed manifolds defined
by~\eqref{eq:Anfixedpts} in more detail. We will again determine the
exact  R-charges by
$a$--maximization~\cite{Intriligator:2003jj}\footnote{A related
discussion appears in~\cite{Franco:2003ja,Franco:2003ea}}. The
main difference with the above approach is that we do not need to use
the $SU(2)$ symmetry. Nevertheless, we will still recover the
result~\eqref{eq:Anfixedptssymm}. 

We
can rewrite~\eqref{eq:Anfixedpts} 
\begin{equation}
\begin{split}
\left. \begin{matrix}
\hspace*{7mm}R_{\Phi_i} + R_{A_i} + R_{B_i} -2=0 \\
R_{\Phi_i}+R_{A_{i-1}} + R_{B_{i-1}} -2  =0
\end{matrix} \right\}~&~\forall~ i|m_i=0,  \\
\left. \begin{matrix}
\hspace*{7mm} R_{A_i}  + R_{B_i} -1= 0 \\ 
R_{A_{i-1}}+R_{B_{i-1}}-1 =0
\end{matrix} \right\} ~&~\forall~ i|m_i\neq 0.
\end{split} \label{eq:AnRconstraints}
\end{equation}
Next we compute the $a$--charge, which is given
by~\cite{Anselmi:1998am,Anselmi:1998ys}
\begin{equation}
a = \frac{3}{32} \left( 3 \Tr R^3 - \Tr R\right),
\end{equation}
where the traces are performed over all fermions in the
theory. Because some adjoints are integrated out, we need to modify
the formula~\eqref{eq:acharge} accordingly. For
the $\widehat{A}_{n-1}$ theories we find that
\begin{equation}
\begin{split}
a = & \frac{3}{32} \left[ 2nN^2 
+ N^2 \sum_{i|m_i=0} \left( 3 (R_{\Phi_i}-1)^3 - (R_{\Phi_i}-1)
\right) \right.
\\
& \left. \hspace{1cm} +  N^2 \sum_i \left( 3 (R_{A_i}-1)^3 - (R_{A_i}-1) 
+ 3 (R_{B_i}-1)^3 - (R_{B_i}-1)\right)
\right].
\end{split} \label{eq:Anacharge}
\end{equation}

Let us first examine $\widehat{A}_{1}$ with $m_1,m_2\neq 0$. Then both
$\Phi_{1,2}$ are integrated out and~\eqref{eq:AnRconstraints} become
\begin{equation}
R_{A_1}  + R_{B_1} -1 =
R_{A_2} +R_{B_2}-1 =0. 
\end{equation}
The $a$--charge~\eqref{eq:Anacharge} is
\begin{equation}
a =  \frac{3}{32} \left[ 2nN^2 
+  N^2  \left( 3 \left[ (R_{A_1}-1)^3  -R_{A_1}^3 \right]
+ 3 \left[ (R_{A_2}-1)^3  -R_{A_2}^3 \right] +2
\right)\right]. \label{eq:Aoneacharge}
\end{equation}
The only extremum is a maximum at the point 
\begin{equation}
R_{A_1} =R_{A_2}=R_{B_1}=R_{B_2}=\frac{1}{2}. \label{eq:A1max}
\end{equation}

The extremization equations for general $\widehat{A}_{n-1}$ with all
$m_i\neq 0$ will take 
the same form as the ones derived
from~\eqref{eq:Aoneacharge}. Therefore the R--charges at generic fixed
points when all $m_i\neq 0$ always 
satisfy 
\begin{equation}
R_{A_i} =R_{B_i}=\frac{1}{2},~~R_{\Phi_i}=1. \label{eq:AnRcharges}
\end{equation}

This computation justifies the R--charge assignments
of~\cite{Bergman:2001qi}. 
In fact, since~\eqref{eq:AnRconstraints} imply that
\begin{equation}
R_{A_i} + R_{B_i} = R_{A_{i+1}} + R_{B_{i+1}},~~m_i=0,
\end{equation}
the cyclicity of the quiver will result in the R--charge
assignments~\eqref{eq:AnRcharges} even when $\mu<n$ of the
$m_i=0$. Since the 
fermionic components of $\Phi_i$ drop out of the R--charge traces, 
the computation of~\cite{Bergman:2001qi} that 
$\tfrac{c_{\text{\tiny IR}}}{c_{\text{\tiny UV}}} = \tfrac{27}{32}$
is valid for all  flows away from $\widehat{A}_{n-1}$ generated by
masses~\eqref{eq:massdef}. This fits well with the supergravity
analysis of~\cite{Corrado:2002wx}.

We note that all of the gauge invariant
operators at the fixed point 
satisfy the unitarity bound $R\geq \tfrac{2}{3}$ and that
$a_{\text{\tiny UV}}> a_{\text{\tiny IR}} 
= \tfrac{27}{32}a_{\text{\tiny UV}}$ so that the $a$--theorem is
satisfied for these flows~\cite{Cachazo:2001sg}. 

\subsection{Moduli space geometry}

We would now like to study the moduli space of the
theory~\eqref{eq:Asuppot}. The F--flatness conditions for arbitrary
$m_i$ are
\begin{equation}
\begin{split}
 \delta_{m_i,0} \lambda_i\left(A_{i} B_{i}-B_{i-1}A_{i-1}\right) &=0,
\\
-\delta_{m_i,0} \lambda_i B_i\Phi_i 
+ \delta_{m_{i-1},0} \lambda_{i+1} \Phi_{i+1} B_i \hspace*{2.5cm} & \\
+ (1-\delta_{m_i,0}) \, h_i\,
\left( B_i A_i B_i-B_iB_{i-1}A_{i-1}\right) \hspace*{.7cm} & \\
+(1-\delta_{m_{i+1},0}) \, h_{i+1}\, 
\left( B_i A_i B_i-A_{i+1} B_{i+1} B_i\right) 
&=0 .
\end{split} \label{eq:AnFflat}
\end{equation}
There is an additional equation arising from $\partial W/\partial B_i$
which is analogous to the second equation above.

The general features of the geometry of the moduli space can be
obtained by studying the $U(1)_{\text{\small{diag.}}}\subset U(N)^n$
degrees of freedom
\begin{equation}
A_i = a_i \, \Bid_{N\times N}+\cdots,
~~B_i = b_i \, \Bid_{N\times N}+\cdots,
~~\Phi_i = \phi_i \, \Bid_{N\times N}+\cdots. \label{eq:Uonedof}
\end{equation}
The F--flatness conditions will result in a moduli space for these
degrees of freedom which we can denote by $\CM$. Then projecting the
F--flatness conditions onto the other components of the Cartan
subalgebra of $U(N)$ will lead to additional copies of $\CM$.
The Weyl group $S_N$ acts on these components, so the full moduli
space is the symmetric product
\begin{equation}
S^N \CM = (\CM)^N/S_N.
\end{equation}
In the following, we will refer to $\CM$ as the ``moduli space.''

We define $\BZ_n$--invariant coordinates by 
\begin{equation}
x=\prod_i a_i, ~~y =\prod_i b_i, ~~z_i = a_i b_i. \label{eq:invcoords}
\end{equation}
Then~\eqref{eq:AnFflat} become
\begin{equation}
\begin{split}
 \delta_{m_i,0} ( z_i- z_{i-1} ) &=0,
\\
-\delta_{m_i,0}  \phi_i  
+ \delta_{m_{i+1},0}  \phi_{i+1}  
+ (1-\delta_{m_i,0}) \,h_i \,
\left( z_i -z_{i-1}\right) & \\
+(1-\delta_{m_{i+1},0}) \, h_{i+1} \,
\left( z_i-z_{i+1} \right) 
&=0 .
\end{split} \label{eq:AnFflatUone}
\end{equation}

We first consider the case
that all $m_i\neq 0$. Then, using~\eqref{eq:hdef}, the
equations~\eqref{eq:AnFflatUone} become a matrix equation 
\begin{equation}
M_{ij} z_j = 0,~~
M_{ij} = (h_i+h_{i+1})\delta_{ij} 
- h_{i+1}\delta_{i+1,j} - h_i \delta_{i-1,j}. \label{eq:Anmatrix}
\end{equation}
We would like to compute the rank of $M$. We find that we can express 
\begin{equation}
\det\, M = (h_1 + h_2 -h_1 - h_2) \,  
\left(\prod_{i=1}^n h_i\right)\left(\sum_{i=1}^n h_i^{-1}\right) =0.
\end{equation}
Since $\det\, M=0$, $M\cdot z=0$ always has non--trivial solutions. We
further note that if $S$ is any $n-1\times n-1$ submatrix of $M$, then
\begin{equation}
\det\, S = f(h_i) \left(\sum_{i=1}^n h_i^{-1}\right).
\end{equation}
Therefore $M$ has rank
$n-1$ when $\sum_i h_i^{-1} \neq 0$ and has rank $n-2$ when 
$\sum_i h_i^{-1} = 0$. 

When $M$ has rank $n-1$, the solutions of~\eqref{eq:Anmatrix} are given
by $z_i = c(z)$ for some function $c(z)\neq 0$. From the homogeneity
of the problem, we should take $c(z)$ to be linear, $c(z)=z$, where we
have absorbed a possible numerical factor into the definition of $z$.  
Then we find that the moduli space is given by 
\begin{equation}
xy=z^n ,\label{eq:Ansingcurve}
\end{equation}
{\it i.e.}, it is the singular $A_{n-1}$ curve in $\BC^3$. 

Recall that the existence of the fixed point 
requires that the F-term coefficients $f_i=0$. However, we are free to
add D--terms, so the singularity of~\eqref{eq:Ansingcurve} can be
resolved, but not deformed by a modification of the complex
structure.  This has an interpretation in the string dual
picture. Evidently the untwisted 3--form flux not only generates a
potential on the probe brane, but it also creates an obstruction to
complex deformation of the ALE space. This is possible, since the
metric on the ALE space is no longer Ricci flat. Therefore there is no
hyperK\"ahler isometry to relate the  resolutions to the  complex
structure deformations.   

When $M$ has rank $n-2$, we again have the solution $z_i =  z$.
However now there are also solutions to the submatrix equation 
$S\cdot z=0$. By operations on the rows of 
$M$, these generate additional solutions to $M\cdot z=0$ of the form
$z_i = \gamma_i \tau(t)$, where  $t$ is an independent complex
variable.    
However, these $z_i$ should scale as $t\sim \langle \Tr \Phi_i\rangle$,
so it is natural to choose $\tau(t)=t$.  Putting these solutions
together, we have $z_i  
= z- \gamma_i \, t$. By a translation in $z$, we can set
$\sum_i\gamma_i=0$.  
Note that $M\cdot \gamma=0$ and $\sum_i\gamma_i=0$ are $n-1$
independent equations for the $n$ $\gamma_i$. 
These have the solution $\gamma_i-\gamma_{i-1}=h_i^{-1}$.
The remaining degree of freedom can be absorbed into a rescaling of
$t$.   

The moduli
space is 
\begin{equation}
\begin{split}
& xy= \prod_{i=1}^n (z -\gamma_i \, t) , \\
& \gamma_i = \sum_{j=1}^{i} h_j^{-1} 
- \frac{1}{n} \sum_{j=1}^n (n-j) h_j^{-1}. 
\end{split}
\label{eq:Andefcurve}
\end{equation}
The moduli space is a deformed $A_{n-1}$ curve fibered over a complex
line parameterized by $t$. This manifold is a  generalized conifold,
and its appearance in this field theory was discussed
by~\cite{Gubser:1998ia,Cachazo:2001gh}.  The parameters $h_i$ determine
the complex structure moduli of this manifold.

Now suppose that $\mu$ of the $m_i$ vanish. The
equations~\eqref{eq:AnFflatUone} now give an equation of the form
\begin{equation}
\widetilde{M}_{ij} Z_j =0, ~~ 
Z_i = (z_1,\ldots,z_{n-\mu},\phi_1,\ldots,\phi_\mu),
\end{equation}
where we have made a relabeling of variables. The matrix
$\widetilde{M}$ has $\det\,\widetilde{M}=0$. Also the submatrix acting
on the $z_i$ subspace has the same form as $M_{ij}$
in~\eqref{eq:Anmatrix}. We find that $\widetilde{M}$ has rank $n-\mu-1$ if
$\sum_i h_i^{-1} \neq 0$ and rank $n-\mu-2$ if 
$\sum_i h_i^{-1} = 0$. In the former 
case, we obtain the resolvable $A_{n-1}$ curve~\eqref{eq:Ansingcurve} as
the moduli space. For the latter, we will find that
\begin{equation}
xy= z^{\mu} \prod_{i=1}^{n-\mu} (z -\gamma_i \,t).
\end{equation}
These are generalized conifolds corresponding to partial resolutions
of the $A_{n-1}$ singularities.  Existence of the fixed point rules
out adding F--terms, except perhaps in some very special situations.

\section{The Spectrum of Marginal Operators at the Fixed Points}
\label{sec:marginal}

We would like to further analyze the manifolds of fixed points
discussed in Section~\ref{sec:massdef}. Specifically we would like
elucidate the form of the marginal operators that parameterize the
manifolds of fixed points. 

Motion in the moduli space $\CM_\tau^{(\nu)}$ of gauge couplings is
generated by exactly 
marginal operators corresponding to the differences between the gauge
kinetic  
energies. We are interested in motion on the moduli space of mass
deformations, parameterized by the $h_i(\tau_i)$, so we will study the
theories with all $m_i \neq 0$.   These fixed points
are defined by superpotentials
\begin{equation}
\begin{split}
& W = - \, \Tr  
\sum_{i} \frac{h_i}{2} T_i, \\
& T_i = \left(A_{i} B_{i}-B_{i-1}A_{i-1}\right)^2 . 
\end{split} \label{eq:Afpsuppot}
\end{equation}

Let us first consider the $A_1$ theory. The moduli space is a $\BP^1$
parameterized by homogeneous coordinates $(h_1,h_2)$, as depicted in
Figure~\ref{fig:Aone}. We find that 
we can rewrite~\eqref{eq:Afpsuppot} as
\begin{equation}
\begin{split}
& W = - \frac{1}{4} \, \Tr  
\Bigl[ (h_1-h_2) \left( A_1 B_1 B_2 A_2 
- A_2 B_2 B_1 A_1 \right) \\
& \hspace*{3.5 cm}  
-(h_1+h_2) \Bigl((A_1 B_1-B_2A_2)^2 
+  (A_2 B_2-B_1A_1)^2 \Bigr) \Bigr].
\end{split} \label{eq:Aonefpsuppot}
\end{equation}

The case $m_1+m_2=0$ was studied in~\cite{Klebanov:1998hh}. They
argued that a $\BZ_2$ symmetry could be used to fix the gauge
couplings to be equal along the flow to the fixed point,
$\tau_1=\tau_2$. Therefore we also have $h_1+h_2=0$. The
superpotential~\eqref{eq:Aonefpsuppot} reduces to
\begin{equation}
W =\lkw \,\Wkw = \lkw \,\Tr  
\left( A_1 B_1 B_2 A_2 
- A_2 B_2 B_1 A_1 \right),
\end{equation}
which agrees with their superpotential after one notes the change of
notation  $A_2^{\text{\tiny here}}=B_2^{\text{\tiny KW}}$, 
$B_2^{\text{\tiny here}}=A_2^{\text{\tiny KW}}$. Note that $m_1+m_2=0$
corresponds to the point $(1,-1)$ on $\BP^1$, which we referred to as
the KW point in subsection~\ref{sec:illustrate}. It was noted
in~\cite{Klebanov:1998hh} that this point has an $SU(2)\times SU(2)$
global symmetry. In our notation, each factor acts independently on
the doublets $(A_1, B_2)$ and $(A_2, B_1)$. The $SU(2)$ at a generic
point of $\BP^1$ which was inherited from the $\CN=2$ R--symmetry is
the diagonal subgroup of this enhanced symmetry. Furthermore,
\cite{Klebanov:1998hh} explained how this theory described $N$
D3-branes on the conifold. The result~\eqref{eq:Andefcurve} reduces
appropriately.

The point on $\BP^1$ which is antipodal to the KW point is
$m_1=m_2$, which is the PW point. This point also has a $\BZ_2$
symmetry which can be used to set $\tau_1=\tau_2$, so $h_1=h_2$ as
well. Then \eqref{eq:Aonefpsuppot} reduces to 
\begin{equation}
W =\lpw \, \Wpw = \lpw \,\Tr  
\Bigl[ (A_1 B_1-B_2A_2)^2 + (A_2 B_2-B_1A_1)^2  \Bigr].
\end{equation}
In~\cite{Corrado:2002wx} it was argued that this point describes $N$
D3-branes at a cone over a $\BZ_2$ orbifold of the fixed point
solution of~\cite{Pilch:2000ej,Pilch:2000fu}. This solution has
non--zero complex 3--form flux, which generates a potential for a
D3-brane probe~\cite{Johnson:2000ic}. The minimum of this potential
determines the moduli space to be a singular $A_1$ ALE space, in
agreement with the analysis leading to~\eqref{eq:Ansingcurve}.

Now the whole $\BP^1$ of fixed point theories can be described by
rewriting~\eqref{eq:Aonefpsuppot} as
\begin{equation}
W =\lkw \, \Wkw + \lpw \, \Wpw .
\end{equation}
The $\BP^1$ is recovered as the fixed line of solutions to the
equations
\begin{equation}
\begin{split}
& \gamma_\Wkw(\tau,\lkw,\lpw) +2 =0, \\ 
& \gamma_\Wpw(\tau,\lkw,\lpw) +2 =0 .
\end{split}
\end{equation}
We can take the difference
\begin{equation}
\CO = \frac{1}{2} \left( \Wkw - \Wpw \right) = 2(A_2 B_2-B_1A_1)^2
\end{equation}
to be the exactly marginal operator which generates translations on
the $\BP^1$ manifold of fixed points.

In the $A_{n-1}$ case, we will have a $\BP^{n-1}$ manifold of fixed
points specified by the conditions
\begin{equation}
\gamma_i(\tau_i,h_i) +2 =0 
\end{equation}
on the anomalous dimensions of the operators $T_i$ defined
in~\eqref{eq:Afpsuppot}.  The analog of the PW point is the point 
$m_i=m,\, \forall i$.  Now this point describes $N$
D3-branes at a cone over a $\BZ_n$ orbifold of the fixed point
solution of~\cite{Pilch:2000ej,Pilch:2000fu}. The $\BZ_n$ symmetry can
be used to set the gauge couplings equal $\tau_i = \tau$, so that
$h_i=h,\, \forall i$. The operator defining
this point is  
\begin{equation}
\Wpw = \Tr  \sum_i T_i.
\end{equation}

The equation $\sum_i h_i^{-1}=0$ defines a subvariety of $\BP^{n-1}$
that is isomorphic to 
$\BP^{n-2}\subset \BP^{n-1}$. Therefore the analog of the KW fixed
point is a $\BP^{n-2}$  submanifold. We will refer to this submanifold
as the conifold subspace. Let us consider the perturbation of the
superpotential at a point on the conifold
subspace by the marginal operator
\begin{equation}
\CO = \Tr  \sum_i c_i T_i. \label{eq:genpert}
\end{equation}
This perturbation takes us to a new point defined by the
superpotential 
\begin{equation}
W'  = \Tr  \sum_i (h_i +c_i) \, T_i
\equiv \Tr  \sum_i h'_i \, T_i.
\end{equation}

In general the point specified by the $h'_i$ is no longer on the
hyperplane $\sum_i h_i^{-1}=0$. The condition that this point is still on
the conifold subspace is 
\begin{equation}
0=\sum_i (h'_i)^{-1} = \sum_i \frac{1}{h_i+c_i } .  \label{eq:pintersect} 
\end{equation}
This has the solution
\begin{equation}
c_i = -\frac{h_i\gamma_i}{1+h_i \gamma_i} ,~~\sum_i \gamma_i =0.
\label{eq:concoeff}
\end{equation}
Then~\eqref{eq:genpert} takes the point $h_i^{-1}$ to the point 
$(h'_i)^{-1}=h_i^{-1} +\gamma_i$.   

A basis for general perturbations of a fixed point can be chosen as
the operators
\begin{equation}
\CO_i = \Tr  \left( T_i - T_{i-1}\right).
\end{equation}
As we found above, a generic perturbation~\eqref{eq:genpert} will move
a point on the conifold subspace off into the $\BP^{n-1}$. Perturbations
which move a point to another point on the conifold subspace depend on
the initial condition according to~\eqref{eq:concoeff}. Therefore a
basis for perturbations within the conifold subspace is
\begin{equation}
\CO_i(\gamma) = \Tr  \left[  
-\frac{h_i\gamma}{1+h_i \gamma} T_i
+\frac{h_{i-1}\gamma}{1-h_{i-1} \gamma} T_{i-1} \right]. 
\label{eq:conifoldconserve}
\end{equation}
These take $(h_1^{-1},\ldots,h_{i-1}^{-1},h_i^{-1},\ldots,h_n^{-1})$
to 
$(h_1^{-1},\ldots,h_{i-1}^{-1}-\gamma,h_i^{-1}+\gamma,\ldots,h_n^{-1})$.
The operators~\eqref{eq:conifoldconserve} form an Abelian group under
compositions $\CO_i(\gamma)\cdot \CO_{i'}(\gamma')$.

\section{Deformation of $\CN=2$ Conformal Theories by General $W(\Phi_i)$}
\label{sec:gendef}

One can also consider deformations by general polynomials 
\begin{equation}
\begin{split}
& W(\Phi_1,\ldots,\Phi_n) = \sum_i W_i(\Phi_i), \\
& W_i(\Phi_i) = \Tr \sum_{r=0}^{k} \frac{g^{(i)}_{r+1}}{r+1} 
\, \Phi_i^{r+1},
\end{split} \label{eq:gendef}
\end{equation}
as in~\cite{Cachazo:2001gh,Cachazo:2001sg,Oh:2001bf}. 
For a generic perturbation, it is not possible to analyze the theory
by integrating out the $\Phi_i$, so one must study the F--flatness
conditions (we specialize to $\widehat{A}_{n-1}$ for convenience)
\begin{equation}
\begin{split}
& \lambda_i\left(A_{i} B_{i}-B_{i-1}A_{i-1}\right)  
+ W_i'(\Phi_i)=0, \\
& - \lambda_i B_i\Phi_i  +  \lambda_{i+1} \Phi_{i+1} B_i =0 ,\\
&  \lambda_i \Phi_i A_i -  \lambda_{i+1}  A_i\Phi_{i+1} =0.
\end{split} \label{eq:genFflat}
\end{equation}
The first equation is consistent with cyclicity for two cases. Either
$\sum_i W_i'(\Phi_i)/\lambda_i=0$ or $\langle\Phi_i\rangle=0$ for all $i$. In
the latter case, we must also demand that $\sum_i g^{(i)}_{1} =0$,
which is just the familiar condition on the F--terms. When
$\sum_i W_i'(\Phi_i)/\lambda_i\neq 0$, the moduli space is an ALE space.

When  $\sum_i W_i'(\Phi_i)/\lambda_i=0$, the second and third
equations require  that $\lambda_i\Phi_i=\Phi$ for all $i$. 

Then the equations for the 
$U(1)_{\text{\small{diag.}}}$ degrees of freedom~\eqref{eq:Uonedof}
become
\begin{equation}
z_i - z_{i+1} +   \frac{1}{\lambda_i}W_i'(t) =0. \label{eq:zgendef}
\end{equation}
We have absorbed the gauge coupling into the
definition of the $\BZ_n$--invariant
coordinates~\eqref{eq:invcoords} and introduced the
coordinate $t=\tfrac{1}{\widetilde{N}}\Tr \Phi$.  The
equations~\eqref{eq:zgendef} can be solved by setting $z_1 = z+c$ and
computing the other $z_i$ by recursion. The moduli space
obtained is~\cite{Cachazo:2001gh,Oh:2001bf} 
\begin{equation}
\begin{split}
& xy=\prod_i (z - \tau_i(t)), \\
& \tau_i(t) = \sum_{j=1}^{i} \frac{W_j'(t)}{\lambda_j}
- \frac{1}{n} \sum_{j=1}^n (n-j)  \frac{W_j'(t)}{\lambda_j}.
\end{split} \label{eq:Angencon}
\end{equation}
The $\tau_i(t)$ are degree $n-1$ polynomials in $t$. This is the  most
general deformation  of the $A_{n-1}$ curve to a generalized
conifold~\cite{KatzMorrison,Cachazo:2001gh}.   

\subsection{Restrictions from conformal invariance}

This computation of the moduli space is not the complete
story. After deforming the $\CN=2$ fixed point by~\eqref{eq:gendef},
the theory should flow to some conformal fixed point in the IR. The
moduli space should then reflect the geometry dual to this IR fixed
point. However, \eqref{eq:gendef} generally contains operators
which are irrelevant at the $\CN=2$ fixed point. 

This issue was addressed in~\cite{Cachazo:2001sg} by an argument that
the operators in~\eqref{eq:gendef} are actually dangerously
irrelevant. In analogy to~\cite{Kutasov:1996ss}, they argue that at
large $g^{(i)}_{k+1}$ there is a fixed point where the operators 
\begin{equation}
W_i(\Phi_i) = \Tr\frac{g^{(i)}_{k+1}}{k+1} \Phi_i^{k+1} 
\label{eq:monodef}
\end{equation}
become marginal. However, we will now present an argument that there
are no dangerously irrelevant operators for $k>2$. 

Assuming that some version of an $a$--theorem for 4D RG flows is true,
a crucial criterion for the
existence of a fixed 
point is that the $a$--theorem is satisfied for flows  generated by
relevant deformations at the candidate fixed
point~\cite{Intriligator:2003mi}.
In~\cite{Cachazo:2001sg} the $a$--charge for the candidate fixed points
generated by the perturbations~\eqref{eq:monodef} was found to be
\begin{equation}
a_k = \frac{27k^2 \widetilde{N}^2 \,  |\Gamma|}{16(k+1)^3} . 
\label{eq:kacharge}
\end{equation}
A class of relevant operators at the $\Tr \Phi^{k+1}$ candidate fixed
points are the operators $\Tr \Phi^{k'+1}$ with $k'<k$. Perturbations
by $\Tr \Phi^{k'+1}$ would drive the theory toward the candidate fixed
point where $\Tr \Phi^{k'+1}$ becomes marginal. However, since 
\begin{equation}
\frac{da_k}{dk} = 
-(27  \widetilde{N}^2\, |\Gamma|) \frac{k(k-2)} {16(k+1)^4}
\end{equation}
is negative definite for $k>2$, the $a_k$ charge for these
candidate fixed points is a strictly decreasing function. Therefore
$a_{k'}> a_k$ whenever $k'<k$ and these flows would always violate
the $a$--theorem. We conclude that $\Tr \Phi^{k+1}$, $k>2$ do {\it not}
generate new
superconformal fixed points. The operators $\Tr \Phi^{k+1}$ are simply
irrelevant when $k>2$.

This result does not contradict the fact that the operators analogous
to~\eqref{eq:gendef} in SQCD with one adjoint are dangerously
irrelevant~\cite{Kutasov:1996ss,Kutasov:2003iy}. If we set all but one
gauge coupling to zero, we obtain a $U(N_c)$ gauge theory with 
$N_f=2N_c$ quarks.  We can apply the analysis of~\cite{Kutasov:2003iy}
to this theory. 
They found that the operators~$\Tr \Phi^{k+1}$ defined new fixed
points for $N_c$ and $N_f$  satisfying $N_c/N_f> x_k$, where for
small $k$,  
\begin{equation}
x_k = \sqrt{\frac{1}{20}\left(\frac{(5k-4)^2}{9}+1\right)}.
\end{equation}
As $x_k>\tfrac{1}{2}$ for $k>2$, these fixed points never exist
for the value $N_f=2N_c$ corresponding to the quiver theories.

As noted in~\cite{Cachazo:2001sg}, the candidate fixed point for the
cubic operator $\Tr\Phi^3$ satisfies 
$a_{k=2} = a_{\text{\tiny free}}$. Since $a_{k=1}<a_{k=2}$, there
is no $a$--theorem  
violation for the deformation of this by a mass term $\Tr \Phi^2$. In
adjoint SQCD, \cite{Kutasov:2003iy} determined that
$x_2=\tfrac{1}{2}$, so it is possible that the cubic operator in this
theory is marginally relevant. However,  the present analysis cannot
decisively rule out or prove the existence of these candidate fixed
points.  We note that a perturbative analysis described
in~\cite{Cachazo:2001sg} suggests that $\Phi_i^3$ is marginally
irrelevant for small couplings, but the large coupling behavior is
still unknown. 

\subsection{The IR fixed point moduli spaces revisited}

The above results imply that if we add a potential $W(\Phi_i)$ to the
$\CN=2$ conformal theory, the higher order terms in $W(\Phi_i)$ are
irrelevant. The analysis that lead to~\eqref{eq:Angencon} is strictly
only valid in the UV, where the irrelevant operators in $W(\Phi_i)$
are still important. At the fixed point in the IR, the effective
$W(\Phi_i)$ will only contain marginal operators, which can come from
operators that were relevant in the UV. If $W(\Phi_i)$ does not
contain any relevant operators, it is possible that some are generated
at critical points of $W(\Phi_i)$. If they are not, the theory will
flow back to the undeformed $\CN=2$ theory, with its orbifold moduli
space.  

We will consider the case that the $\Phi_i$ are near a
critical point of $W(\Phi_i)$ and compute the relevant
part of $W(\Phi_i)$ at this critical point. This will be
a sum of mass terms for the perturbations around the critical point
and we can compute the moduli spaces reliably using the analysis of
Section~\ref{sec:massdef}. We will then argue that this is consistent
with the validity of~\eqref{eq:Angencon} away from the critical
points. However, it will still be important that $W(\Phi_i)$ contain
mass terms in the UV in order to remove the orbifold singularities in
the moduli space. 

Suppose that $\phi_i$ is a critical point of $W_i(\Phi_i)$. Applying
the F--flatness conditions~\eqref{eq:genFflat}, we find that
$\phi_i=\phi, \, \forall i$. Clearly this is easiest to accomplish if
$W_i(\phi)/\lambda_i=W(\phi), \,\forall i$, but we will not require
this. We now 
set $\Phi_i = \phi + \widetilde{\Phi}_i/\lambda_i$ and expand to
quadratic order  
\begin{equation}
W_i(\Phi_i) =  W_i(\phi) 
+ \frac{1}{2} W_i^{''}(\phi) 
\left(\frac{\widetilde{\Phi}_i}{\lambda_i}\right)^2 + \cdots.
\end{equation}
The F--flatness conditions are now
\begin{equation}
\begin{split}
& \left(A_{i} B_{i}-B_{i-1}A_{i-1}\right)  
+ \frac{W_i^{''}(\phi)}{\lambda_i^2} \widetilde{\Phi}_i=0, \\
& -  B_i\widetilde{\Phi}_i  
+   \widetilde{\Phi}_{i+1} B_i =0 ,\\
&   \widetilde{\Phi}_i A_i 
-    A_i\widetilde{\Phi}_{i+1} =0.
\end{split} \label{eq:genFflattilde}
\end{equation}

If $\sum_i W_i^{''}(\phi)/\lambda_i^2\neq 0$, then the first equation
of~\eqref{eq:genFflattilde} is consistent with cyclicity only for
$\widetilde{\Phi}_i=0$. We will recover a moduli space which is just
the ALE space. 

If $\sum_i W_i^{''}(\phi)/\lambda_i^2= 0$, then the second and third equations
of~\eqref{eq:genFflattilde} require that
$\lambda_i\widetilde{\Phi}_i=\widetilde{\Phi}, \, \forall i$. Now the solution
presented in Section~\ref{sec:massdef} involves a coordinate $t$ that
is related to the VEVs of $\langle \Tr\Phi \rangle$. Because we are
close to the critical point, we set $t\equiv \phi + \tilde{t}$, where
$\tilde{t}\equiv \tfrac{1}{\widetilde{N}} \langle\Tr \tilde{\Phi}_i
\rangle$. The moduli space is then 
\begin{equation}
\begin{split}
& xy=\prod_i (z - \tau_i(\tilde{t},\phi)), \\
& \tau_i(\tilde{t},\phi) = \tilde{t}
\left[\sum_{j=1}^{i} \frac{W_j^{''}(\phi)}{\lambda_j^2}
- \frac{1}{n} \sum_{j=1}^n (n-j)  
\frac{W_j^{''}(\phi)}{\lambda_j^2}\right].
\end{split} \label{eq:critmod}
\end{equation}

The result~\eqref{eq:critmod} agrees with the expansion
of~\eqref{eq:Angencon} around the critical points, $t =  \phi +
\tilde{t}$,  of the $W_i(t)$.  It is natural then to assume
that~\eqref{eq:Angencon} is the correct result away from the critical
points. Evidently, when the irrelevant $\Phi_i^{k+1}$ decouple as the
theory flows to the IR, the data of their coupling constants is
reflected in the running of the couplings of the less irrelevant
operators. From~\eqref{eq:Angencon} we conclude that the theory flows
to a fixed point generated by an effective mass $W_i'(t)$ for the
$\Phi_i$.  

In order for~\eqref{eq:Angencon} not to have orbifold singularities, it
is crucial that $W(\Phi_i)$ actually contains non--zero bare mass
terms for all of the $\Phi_i$. Suppose that $W_i(\Phi_i)$ does not
have a mass term. Then the $i^{\text{th}}$ sector will flow to the
least irrelevant monomial $W_i(\Phi_i) \sim \Phi_i^{k+1}$. This
operator is still irrelevant if $k>2$, but it is possible that
relevant operators are generated at a critical point. However, the
only critical point for this potential is $\phi_i=0$, so
$W_i^{''}(\phi_i)=0$ (no relevant operator is generated). Then at
least one of the differences $\tau_i - \tau_j$ in~\eqref{eq:critmod}
will vanish. Correspondingly, \eqref{eq:critmod} only corresponds to a
partial resolution of the $A_{n-1}$ orbifold singularity.  In some
cases, it may be possible to add F--terms to generate a critical point
for $W_i(\Phi_i)$ at $\phi_i = \phi \neq 0$. 

\section{Nonconformal Theories and Quantum Moduli Spaces}
\label{sec:nonconformal}

A generic $\CN=2$ quiver gauge theory will not be
conformal. Nonconformal theories can be obtained from the $\CN=2$
theory on D3--branes at an orbifold by adding fractional
branes. Deformations of these theories by adjoint masses lead to
$\CN=1$ field theories that illustrate many interesting
features~\cite{Gubser:1998fp,Klebanov:1999rd,Klebanov:2000nc,%
Klebanov:2000hb,Cachazo:2001gh,Cachazo:2001sg}.  

There are two effects we want to analyze. Firstly, since the coupling
constants in these theories are running, it is common that some of the
gauge groups will become strongly coupled in the IR. At these points,
a better description of the theory is in terms of a Seiberg dual
theory~\cite{Seiberg:1995pq}. As one continues to flow into the IR,
the theory can undergo repeated Seiberg dualities, leading to a
duality cascade~\cite{Klebanov:2000hb}. We therefore analyze the
effect of performing a Seiberg duality at a node of the quiver.  

Secondly, these $\CN=1$ theories can have quantum corrections to the
low--energy superpotential~\cite{Affleck:1984mk}, as well as
independent corrections to the classical moduli
space~\cite{Seiberg:1994bz}. We study the moduli space seen by a
D3--brane probe in the dual geometry. 

%------------------------------------------------------------------------------------------------

\subsection{The structure of the effective superpotential}

The $\CN=1$ superpotential in the $A_{n-1}$ theories takes the form
\begin{equation}
\begin{split}
 W& =- \Tr  
\sum_{i} \frac{h_i}{2}  \left(A_{i} B_{i}-B_{i-1}A_{i-1}\right)^2 \\
&= \sum_i \Bigl( 
h_i\, \Tr A_{i} B_{i} B_{i-1}A_{i-1} 
-\frac{h_i+h_{i+1}}{2} \Tr(A_i B_i)^2  \Bigr).
\end{split} \label{eq:UVsuperpot}
\end{equation}
We will take this theory as the UV completion of the IR physics that
we study below, even though~\eqref{eq:UVsuperpot} can be obtained by
deformation of an $\CN=2$ theory. Therefore the couplings $h_i$ are
the fundamental quantities and we will no longer refer to masses
$m_i$.  By analogous arguments to those in section~\ref{sec:massdef},
the moduli space of the theory with $\sum_i h_i^{-1}\neq 0$ is an ALE
space, while that for $\sum_i h_i^{-1}= 0$ is a generalized conifold.

The superpotential~\eqref{eq:UVsuperpot} receives  perturbative
wavefunction renormalizations, which can be 
understood as the running of the coupling constants $h_i$. There are also 
additional nonperturbative corrections allowed. These 
are constrained by the global symmetries and holomorphy.  To determine
the possible corrections, we need to determine what holomorphic
invariants exist. 
These theories have an $SU(2)$ global symmetry that is such that the
products $A_i B_i$ or $B_i A_i$ are 
invariant. Furthermore, there are several $U(1)$s, including a 
nonanomalous baryon $U(1)_B$ and the associated anomalous flavor
symmetry $U(1)_F$. A convenient normalization for these charges is
presented in Table~\ref{tab:reps}. In addition, there is an anomalous
axial symmetry $U(1)_A$. The nonanomalous R--symmetry is a linear
combination of $U(1)_F$ and $U(1)_A$.
\begin{table}[htb]
\begin{center}
\begin{tabular}{|c|c|c|c|c|c|}
\hline
& $U(N_i)$ & $U(N_{i+1})$ & $U(1)_B$ & $U(1)_F$ & $U(1)_A$ \\
\hline
$A_i$ & $\mathbf{N}_i$ & $\overline{\mathbf{N}}_{i+1}$ &
$\tfrac{1}{n\prod_i N_i}$
&  $\tfrac{1}{n\prod_i N_i}$  & 0 \\ 
\hline
$B_i$ & $\overline{\mathbf{N}}_i$ & $\mathbf{N}_{i+1}$ & 
-$\tfrac{1}{n\prod_i N_i}$
&  $\tfrac{1}{n\prod_i N_i}$  & 0 \\ 
\hline
$\Lambda_i^{b_0^{(i)}}$ & & & 0 & 
$(N_{i-1}+N_{i+1})\left(\tfrac{1}{n\prod_i N_i}-1\right)$ 
& $2N_i-N_{i-1}-N_{i+1}$ \\
\hline
$h_i$ & & & 0 & $-\tfrac{4}{n\prod_i N_i}$ & $0$ \\
\hline
\end{tabular}
\end{center}
\caption{The gauge and global symmetry representations of the fields
and couplings.} 
\label{tab:reps}
\end{table}

The holomorphic invariants are of the form
\begin{equation}
I_{\zeta_i\xi_i\chi_i} 
= \Tr \prod_i h_i^{\zeta_i} 
\left(\Lambda_i^{b_0^{(i)}}\right)^{\xi_i} 
\left(A_{i} B_{i}\right)^{\chi_i} , \label{eq:inv}
\end{equation}
where the exponents $\zeta_i$, $\xi_i$, and $\chi_i$ satisfy
\begin{equation}
\begin{split}
& 2\xi_i - \xi_{i-1} - \xi_{i+1} =0, \\
& 4\zeta_i + (N_{i-1}+N_{i+1})\left( n\prod_i N_i -1\right)\xi_i
- 2 \chi_i =0. 
\end{split}
\end{equation}
We should also include invariants which differ from~\eqref{eq:inv} by
inequivalent permutations of the fields. 
The superpotential~\eqref{eq:UVsuperpot} will be renormalized to be of
the form 
\begin{equation}
\begin{split}
 W  = \sum_i \Bigl[ &
h_i\, \left[\Tr A_{i} B_{i} B_{i-1}A_{i-1}\right] \,
F_i(I_{\zeta_i\xi_i\chi_i} ) \\
& - \frac{h_i+h_{i+1}}{2}\left[\Tr(A_i B_i)^2\right]  \, 
G_i(I_{\zeta_i\xi_i\chi_i} )  \Bigr],
\end{split} \label{eq:Wrenorm}
\end{equation}
where the $F_i$ and $G_i$ are some undetermined functions of the 
invariants~\eqref{eq:inv}. 

\subsection{Seiberg Duality}

We want to compute the effect of Seiberg dualizing the group
$U(N_1)$. Therefore we assume that $b_0^{(1)}= 3N_1 - N_n - N_2>0$, so
that $SU(N_1)$ will confine at some scale $\Lambda_1$.  
The $U(N_1)$--invariant degrees of freedom are
the mesons 
\begin{equation}
Z_{ij} = 
\begin{pmatrix} 
A_n A_1  & A_nB_n \\ B_1 A_1 & B_1 B_n
\end{pmatrix}. \label{eq:Zmesons}
\end{equation}
If we also introduce the $U(N_i)$--invariants $M_i = A_i B_i$, we can
rewrite~\eqref{eq:UVsuperpot} as 
\begin{equation}
\begin{split}
W = & h_1 \, \Tr \left[ Z_{11} Z_{22}
- \frac{1}{2} \left( Z_{12}^2 + Z_{21}^2\right) 
-\frac{h_n}{2h_1} \left( Z_{12} - M_{n-1} \right)^2 \right. \\
& \hspace*{2cm} \left. 
- \frac{h_2}{2h_1} \left( Z_{21} -M_2 \right)^2  + \sum_{i=3}^{n-1}
\frac{h_i}{h_1} \left( M_i - M_{i-1}\right)^2
\right]. 
\end{split} \label{eq:ncsp-mesons}
\end{equation}
For now we will ignore the fact that there are
generally nonperturbative renormalizations of~\eqref{eq:ncsp-mesons},
of the form~\eqref{eq:Wrenorm}.

When $ N_n + N_2>N_1$, this theory has a Seiberg dual. The fields
$(A_1,B_1)$, $(A_n,B_n)$ are confined, leaving the mesons $Z_{ij}$ as
low--energy degrees of freedom. The confining gauge group $U(N_1)$ is
replaced by $U(N_n + N_2-N_1)$ and additional degrees of freedom
$(a_1,b_1)$, $(a_n,b_n)$ are added. These fields are in the
bifundamental representations of   $U(N_n + N_2-N_1)\times U(N_2)$ and
$U(N_n)\times U(N_n + N_2-N_1)$, respectively.  We can assemble these
into 2--vectors as $\tilde{q}_i = (a_1,b_n)$, $q_i = (a_n,b_1)$. Then
the superpotential of the dual theory is 
\begin{equation}
\begin{split}
\widetilde{W} = & 
\Tr \left[ y\, Z_{ij}\, q_i \tilde{q}_j 
+ h_1 \,  \left( Z_{11} Z_{22}
- \frac{1}{2} \left( Z_{12}^2 + Z_{21}^2\right) 
-\frac{h_n}{2h_1} \left( Z_{12} - M_{n-1} \right)^2 \right. \right. \\
& \hspace*{4.5cm} \left. \left.
- \frac{h_2}{2h_1} \left( Z_{21} -M_2 \right)^2 \right) \right]+ \cdots.
\end{split} \label{eq:dualsp}
\end{equation}

As the mesons $Z_{ij}$ are massive, they can be integrated out,
leaving the superpotential 
\begin{equation}
\begin{split}
\widetilde{W} = & -\frac{y^2}{h_1} \Tr \left[  X_{11} X_{22}
- \frac{1}{2} \left( X_{12}^2 + X_{21}^2\right) 
-\frac{\tilde{h}_n}{2\tilde{h}_1} \left( X_{12} 
- \widetilde{M}_{n-1} \right)^2 \right. \\
& \hspace*{2cm} \left. 
- \frac{\tilde{h}_2}{2\tilde{h}_1} \left( X_{21} 
-\widetilde{M}_2 \right)^2 \right]
+ \cdots,
\end{split} \label{eq:WdualIO}
\end{equation}
where the new mesons are
\begin{equation}
X_{ij} = 
\begin{pmatrix} 
a_n a_1  & a_nb_n \\ b_1 a_1 & b_1 b_n
\end{pmatrix}, 
~~\widetilde{M}_2=\frac{h_1}{y}M_2,
~~ \widetilde{M}_{n-1}=\frac{h_1}{y}M_{n-1}. 
\end{equation}
This has the same form as the original
superpotential~\eqref{eq:ncsp-mesons}, except that the coupling
constants are shifted as 
\begin{equation}
\tilde{h}_1 = - \frac{y^2}{h_1}, 
~~\tilde{h}_2 = \frac{y^2 h_2}{h_1(h_1+h_2)}, 
~~\tilde{h}_n = \frac{y^2 h_2}{h_1(h_1+h_n)}, 
~~\text{other}~\tilde{h}_i=h_i. \label{eq:sdshift}
\end{equation}
In this equation, all couplings are meant to be defined at the scale
at which the Seiberg duality is performed. 

The case that $n=2$ is slightly different, as $Z_{12}$ and $Z_{21}$
mix. We find that the dual superpotential is  
\begin{equation}
\widetilde{W} =  
-\frac{y^2}{h_1} \Tr \left[  X_{11} X_{22}
- \frac{1}{2} \left(1+ \frac{\tilde{h}_1}{\tilde{h}_2}\right) 
\left( X_{12}^2 + X_{21}^2\right) 
+\frac{\tilde{h}_1}{\tilde{h}_2} X_{12} X_{21} \right], \label{eq:Aonedual}
\end{equation}
where the shift is now
\begin{equation}
\tilde{h}_1 = -\frac{y^2}{h_1}, ~~\tilde{h}_2 = \frac{y^2(2h_1+h_2)}{h_1^2} .
\end{equation}

In the theory studied in~\cite{Klebanov:2000hb}, the bare couplings
satisfy $h_1+h_2=0$. Therefore the theory has an additional $SU(2)$
global symmetry in the UV, as reviewed in
Section~\ref{sec:marginal}. Nonperturbative corrections will not 
generate terms that break this symmetry, so   $h_1+h_2=0$ at all
scales.  As only the ratio $h_1/h_2$ now appears in the
superpotential, the dual 
superpotential~\eqref{eq:Aonedual} is just a rescaling of the original
one~\cite{Strassler:TASI2003}. Therefore the theory
of~\cite{Klebanov:2000hb} can be said to be self--dual, at least in
this sense, under Seiberg
duality~\cite{Intriligator:1995id,Leigh:1995ep}. In general,
especially for $n>2$, 
there are families of 
quartic operators related by Seiberg duality. 

After one Seiberg duality, the theory will continue into the IR until
the next group confines. At that point, so long as
$N_{i-1}+N_{i+1}-N_i>0$, it is possible to Seiberg dualize at the new
confining node. We can then follow the theory to the next confining
scale and repeat the process. This is a duality cascade. A difference
from $A_1$ to $A_{n-1}$ is that some $SU(N_i)$ factors cannot be
Seiberg dualized because $N_{i-1}+N_{i+1}-N_i<0$. Also, if 
$3N_i -N_{i-1}-N_{i+1}<0$, the group $SU(N_i)$ will be IR free.   

Nevertheless, if the $N_i$ are large enough, there will be a duality
cascade over a large range of scales. It is interesting to examine the
result of a large number of Seiberg dualities. For simplicity, we will
consider $A_1$ with no restriction on the bare $h_1+h_2$. Seiberg
dualities will cycle from node to node, so the result
$h^{(s)}_1/h^{(s)}_2$ of performing $s$ Seiberg dualities will depend
on whether $s$ is odd or even. If we begin at node~1, then
we find that 
\begin{equation}
\begin{split}
s=2p+1,\qquad & \frac{h^{(2p+1)}_1}{h^{(2p+1)}_2}
= \frac{h_2-(2p+1)(h_1+h_2)}{h_1+(2p+1)(h_1+h_2)}, \\  
s=2p, \qquad & \frac{h^{(2p)}_1}{h^{(2p)}_2}
= \frac{h_1+2p(h_1+h_2)}{h_2-2p(h_1+h_2)}.
\end{split}
\end{equation}
The behavior for large $s$ is independent of $h_1,h_2$, namely 
$h^{(s)}_1/h^{(s)}_2\rightarrow -1$! We also find that 
$(h^{(s)}_1)^{-1}+(h^{(s)}_2)^{-1}\rightarrow 0$ for large
$s$. Apparently the duality cascade takes the theory to a line of
couplings describing the theory of~\cite{Klebanov:2000hb}.  In the
general case, it appears that we also find that
$\lim_{s\rightarrow 0}\sum_i(h^{(s)}_i)^{-1}\rightarrow 0$.  Even
though the original $h_i$ 
are scale--dependent, it turns out that $\lim_{s\rightarrow
0}\sum_i(h^{(s)}_i)^{-1}$ is not. 

However, the above analysis of the quartic superpotential is
incomplete, due to the nonperturbative 
corrections to~\eqref{eq:ncsp-mesons}. Generally the dual
superpotential at the scale $\Lambda_1$ will not be just a quadratic
polynomial in the mesons, so it is incorrect to integrate them
out via the F--flatness conditions.  
Instead of~\eqref{eq:WdualIO}, the dual superpotential should
take the form
\begin{equation}
\begin{split}
 W  = \sum_i \Bigl[ &
\tilde{h}_i\, \left[\Tr a_{i} b_{i} b_{i-1}a_{i-1}\right] \,
\widetilde{F}_i(\widetilde{I}_{\zeta_i\xi_i\chi_i} ) \\
& - \frac{\tilde{h}_i+\tilde{h}_{i+1}}{2}\left[\Tr(a_i b_i)^2\right]  \, 
\widetilde{G}_i(\widetilde{I}_{\zeta_i\xi_i\chi_i} )  \Bigr].
\end{split} \label{eq:Wdualrenorm}
\end{equation}
Here $(a_i,b_j)$ refer to the dual variables, but if the Seiberg
duality is only performed at node~1, then $(a_i,b_j)=(A_i,B_i)$ when
$i\neq 1,n$. In general the functions $\widetilde{F}_i$ and
$\widetilde{G}_i$ will be different from $F_i$ and $G_i$.
The mesons can still be massive, but they are charged and interact via
various correction terms, so integrating them out to
obtain~\eqref{eq:Wdualrenorm} is hard. 

Nevertheless it is intriguing to consider the possibility that, if all
of these corrections could be taken into account, one would still find
that the net effect of the duality cascade was to take the theory to
the line $\sum_ih_i^{-1}\rightarrow 0$ in coupling space. This would
imply that the theories with different $h_i$ exhibit some sort of
universality in the IR. It is already known that the conformal
theories with different $h_i$ are closely
related~\cite{Corrado:2002wx}.  Perhaps 
information from the gravity dual of this theory can be used to shed
light on the form of the corrections appearing
in~\eqref{eq:ncsp-mesons}, \eqref{eq:Wdualrenorm}, thereby addressing
the question of the true IR behavior of these theories.  

%---------------------------------------------------------------------------------------------------------
\subsection{Infrared Moduli Spaces}

As an example of a theory with a quantum correction to the low--energy
superpotential, we consider  the $A_{n-1}$ generalization of a
theory studied in~\cite{Klebanov:2000hb}. We take an $A_{n-1}$ theory
with gauge group $G=U(N+1)\times  U(1)_2\times \cdots \times U(1)_n$.  
This theory describes a single D3-brane probe in the
background of $N$ fractional branes on one of the homology cycles on
the $A_{n-1}$ fiber of a generalized conifold. In the IR, the interacting
gauge theory is 
$SU(N+1)\subset G$ with 2 flavors. The beta function coefficient is 
$b_0^{(1)} = 2N$, so the interacting theory is asymptotically free and will
confine below some scale $\Lambda_1$. 

Below the scale $\Lambda_1$, we should introduce the
$U(N+1)$--invariant degrees of 
freedom $Z_{ij}$ defined in~\eqref{eq:Zmesons}. The remaining degrees
of freedom can be taken to be the mesons $M_i = A_iB_i$ and the
``baryonic'' operators 
\begin{equation}
x = A_n A_1 \cdots A_{n-1},~~y = B_{n-1}\cdots B_1 B_n.
\label{eq:xandy}
\end{equation}

In terms of these variables, the UV superpotential
is given by~\eqref{eq:ncsp-mesons}. We will call this superpotential
$W_{\text{probe}}$.    
Note that $Z_{11}$ and $Z_{22}$ are charged under 
$U(1)_n\times U(1)_2$, but the combination $Z_{11}Z_{22}$ that enters the
superpotential~\eqref{eq:ncsp-mesons} is invariant. The D--flatness
conditions for these $U(1)s$ are 
\begin{equation}
\begin{split} 
& |Z_{11}|^2 + |B_2|^2 =  |Z_{22}|^2 + |A_2|^2 , \\
& |Z_{11}|^2 + |B_{n-1}|^2 =  |Z_{22}|^2 + |A_{n-1}|^2 .
\end{split}  \label{eq:DflatA3}
\end{equation}
Together with the other D--flatness conditions, we see that these are
a ``solution'' to   
\begin{equation}
x y = Z_{11} Z_{22} M_2 \cdots M_{n-1}. \label{eq:A2oneoneid}
\end{equation}

This theory also generates a dynamical
superpotential~\cite{Affleck:1984mk} in the IR,
\begin{equation}
W_{ADS} = (N-1) \left( \frac{2\Lambda^{3N+1}}{\det Z}\right)^{1/(N-1)},
\label{eq:ADSsp}
\end{equation}
so the complete low--energy effective superpotential is
\begin{equation}
W_{\text{eff}} = W_{\text{probe}} + W_{ADS} .
\end{equation}
Let us consider the F--flatness condition for $Z_{11}$. We find that
\begin{equation}
h_1 \, Z_{22} 
- \left(\frac{2\Lambda^{3N+1}}{(\det Z)^N} \right)^{1/(N-1)} \, Z_{22} 
=0, \label{eq:corrFflat}
\end{equation}
with similar equations for the other $Z_{ij}$. Together these imply that
\begin{equation}
\det Z = c, ~~~~
c = \left( 
\frac{\left(2\Lambda^{3N+1}\right)^{1/(N-1)}}{h_1}
\right)^{N/(N-1)}
\label{eq:Zcdefcon}
\end{equation}
and
\begin{equation}
M\cdot 
\begin{pmatrix} 
Z_{12} \\ Z_{21} \\  M_2 \\ \vdots \\ M_{n-1}
\end{pmatrix} =0, \label{eq:MdotZM}
\end{equation}
where $M$ is of the same form as in~\eqref{eq:Anmatrix}. 

If 
$\sum_i h_i^{-1}\neq 0$, then the solution to~\eqref{eq:MdotZM} is
\begin{equation}
Z_{12} = Z_{21} = M_2  = \cdots = M_{n-1} =z.
\end{equation}
Then~\eqref{eq:Zcdefcon} can be solved for the product
\begin{equation}
Z_{11} Z_{22} = z^2+c .
\end{equation}
Applying these to the identity~\eqref{eq:A2oneoneid}, we find the
moduli space
\begin{equation}
x y =  z^{n-2}(z^2+c). 
\end{equation}
Therefore the $A_{n-1}$ singularity has been partially resolved by the
dynamically generated superpotential. 

When $\sum_i h_i^{-1}=0$, the solution to~\eqref{eq:MdotZM} is 
\begin{equation}
\begin{split}
& Z_{12}=z-\gamma_1 t, ~~ Z_{21} = z-\gamma_2 t,~~M_i =z-\gamma_{i+1} t, \\
& \sum_i \gamma_i=0.
\end{split}
\end{equation}
Rewriting~\eqref{eq:A2oneoneid},  we find
that
\begin{equation}
\begin{split}
xy &=\Bigl[ (z-\gamma_1 t)(z-\gamma_2 t) +c\Bigr] 
\prod_{i=3}^n (z-\gamma_i t) \\
&=  (z-\tau_+(t))(z-\tau_-(t))\prod_{i=3}^n (z-\gamma_i t), \\
&\tau_{\pm}(t) = \tfrac{1}{2}(\gamma_1+\gamma_2) t 
\pm \tfrac{1}{2} \sqrt{(\gamma_1-\gamma_2)^2 t^2 -4c}.
\end{split} \label{eq:A2monodromic}
\end{equation}
This is a generalized conifold which is a monodromic
fibration~\cite{Cachazo:2001gh} because
of the square--root branch cut in $\tau_{\pm}(t)$.  The only
resolvable 2-cycle in the geometry is the one which is wrapped by the
fractional branes.  

A slight generalization is obtained by wrapping fractional branes
around non--adjacent cycles. For example, if we consider $A_3$ with
branes wrapping the cycles corresponding to the first and third nodes,
we obtain a theory with superpotential  
\begin{equation}
\begin{split}
W= &  \Tr \left[ 
h_1 \left( -2 Z_{11}Z_{22} + Z_{12}^2+Z_{21}^2\right)
+h_3 \left( -2 Y_{11}Y_{22} + Y_{12}^2+Y_{21}^2\right)
\right. \\
& \hspace*{2cm} \left. + h_2 \left( Z_{21}-Y_{12}\right)^2
+ h_4 \left( Z_{12}-Y_{21}\right)^2 \right].
\end{split}
\end{equation}
The mesons $Z_{ij}$ and $Y_{ij}$ are defined in an obvious manner
following the conventions of~\eqref{eq:Zmesons}. Also defining the
variables~\eqref{eq:xandy}, the D--terms yield the constraint 
\begin{equation}
x y = Z_{11}  Z_{22} Y_{11} Y_{22}. \label{eq:Afourconst}
\end{equation}

This theory will develop a dynamical superpotential~\eqref{eq:ADSsp}
independently for for the first and third nodes.  The F--flatness
conditions for the resulting low--energy effective superpotential
imply that 
\begin{equation}
\begin{split}
& \det Z = c_1,~~ \det Y= c_3 , \\
& M \cdot 
\begin{pmatrix} Z_{12} \\ Z_{21} \\ Y_{12} \\ Y_{21} \end{pmatrix}
= 0. 
\end{split}
\end{equation}
These can be solved in the usual manner. When $\sum_i h_i^{-1}\neq 0$,
we recover the ALE space 
\begin{equation}
xy = (z^2+c_1)(z^2+c_3).
\end{equation}
When $\sum_i h_i^{-1}= 0$ we find a generalized conifold that is a
monodromic fibration 
\begin{equation}
xy =  \Bigl(z-\tau_{1,+}(t)\Bigr)\Bigl(z-\tau_{1,-}(t)\Bigr)
\Bigl(z-\tau_{3,+}(t)\Bigr)\Bigl(z-\tau_{3,-}(t)\Bigr), 
\end{equation}
where the $\tau_{\alpha,\pm}$ have square root branch cuts as
in~\eqref{eq:A2monodromic}. The generalization to $A_{n-1}$ is
obvious. The only resolvable 2--cycles are the ones which were
originally wrapped by fractional branes. 

These are essentially the only simple examples. In more general cases,
one cannot ignore the fact that fields carry charges under more than
one nonabelian gauge group. This leads to nonperturbative corrections
to the low--energy superpotential~\eqref{eq:Wrenorm}. We will only
attempt to scratch the surface of the corresponding corrections to the
classical geometry.  

Many salient features of the general case are already present in the
example of $A_{n-1}$ with gauge group 
\begin{equation} 
G=U(N+M+1)\times  U(N+1)\times U(1)_3 \cdots \times U(1)_n.
\end{equation}
This theory describes a single probe brane in the presence of
fractional branes wrapping two ``adjacent'' 2-cycles of the ALE
space. It is a generalization of the $A_1$ case
of~\cite{Klebanov:2000hb}. In the IR, the interacting part of the
theory is $SU(N+M)\times SU(N)$ and both gauge groups are
asymptotically free. As we flow to the IR, the group $SU(N+M)$ will
confine first, at the scale $\mu = \Lambda_1$. When $n>2$, it is not
possible to Seiberg dualize $SU(N+M)$.   
Below $\Lambda_1$, the degrees of freedom interacting under $SU(N)$
are the mesons $Z_{ij}$ and the pair $A_2,B_2$. The superpotential in
the UV, $W_0$, has the form~\eqref{eq:ncsp-mesons}.  

If the coupling constant, $g_2$, of $SU(N)$ were zero, the only
correction to the low--energy superpotential would be of ADS--type, 
\begin{equation}
W_{\text{\small eff.}}(g_2=0) 
=  W_0 
+ (M-1) \left( \frac{2\Lambda_1^{b_0^{(1)}}}{\Tr \det Z}\right)^{1/(M-1)}.
\end{equation}
When the $SU(N)$ interaction is turned on, the effective
superpotential is no longer so strongly constrained.  We will not
attempt to determine the precise form of the corrections. On general
grounds, we might expect that the functions appearing
in~\eqref{eq:Wrenorm} include functions of $\Tr \det Z$. One can then
imagine trying to solve the F--flatness conditions by setting  
\begin{equation}
\Tr \det Z  = c,
\end{equation}
for some constant $c$. 
The rest of the F--flatness conditions will lead to a system of $n$
nonlinear equations for the $Z_{ij}$, $M_i$.   Presumably the gauge
theory is smart enough to require that this has a solution, at least
in principle. Then we will find a moduli space of the form
\begin{equation}
xy = P(y,z) ~~\text{or}~~P(y,z,t) .
\end{equation}
where $P(y,\ldots)$ is some function. It is not possible to determine
the dimension of the moduli space without more information.

\section{Discussion}

Our field theory results should be useful in addressing many aspects
of the string theoretic description of these theories. It is known
that the relevant geometry is the generalized conifold, but apart from
the $A_1$ case, where solutions are known for $h_1 = \pm h_2$, no
completely satisfactory explicit classical IIB 
solutions are known. Apart from the case that $h_i =h,\, \forall i$, no 
solutions are known when $\sum_i h_i^{-1} \neq 0$. Explicit, or at least 
approximate, solutions would be necessary
for the computation of field theory correlation functions from the
gravity dual. It is possible that the marginal operators of
Section~\ref{sec:marginal} can be useful for generating new solutions
from the known ones.

We found that when $\sum_i h_i^{-1}\neq 0$, the moduli space of the
$\CN=1$ theory was just the ALE fiber of the generalized
conifold. This has an interpretation as the presence of a potential on
the probe brane which is generated by the 3--form
flux~\cite{Johnson:2000ic}. The probe is transverse to a generalized
conifold, but it is sitting at the minimum of this potential, which is
just the ALE fiber. It would be interesting to verify this in the IIB
duals. 

We also found that the existence of a fixed point demanded that we
could not generally add F--terms to the original $\CN=2$ theory when
deforming by masses leading to  $\sum_i h_i^{-1}\neq 0$. These
F--terms correspond to complex structure deformations of the ALE, so
their absence implies that the 3--form flux presents some sort of
obstruction to complex structure deformation in the dual theory. The
ALE space admits resolutions, in the form of D--terms in the quiver
theory, but since the metric on the ALE is not Ricci--flat, there is
no hyperK\"ahler isometry to relate these to complex structure
deformation. It is important to understand these results in the dual
theory. 

It is also important to firmly resolve the issue of general
deformations of the $\CN=2$ theories. We argued in
Section~\ref{sec:gendef} that dangerously irrelevant operators
$\Phi^{k+1}$ did not exist for $k>2$. This was based on the
$a$--theorem, for which no explicit proof exists. In the absence of
such a proof, it is possible that our use of the $a$--theorem is
invalid. Perhaps  the candidate fixed points generated by
$\Phi^{k+1}$ do exist, and that the flows away from them violate the
$a$--theorem. It could be the case that a different central charge
plays the role of a function that is monotonically decreasing along
the RG flows.  In any case, it is important to better understand the
$a$--theorem in general. It is also interesting to determine whether
$\Phi^3$ is marginally relevant or irrelevant in these theories.  

We used an analysis at the critical points of a general deformation
$W(\Phi_i)$ to argue that the moduli space at the fixed point is given
by~\eqref{eq:Angencon}. This gave a prediction that the coefficient of
the mass terms in  $W(\Phi_i)$ should run to $g_2 \sim W'(t)$ at the
fixed point, after all of the irrelevant operators have dropped out of
the theory. This implies that the coefficients of the $\CN=1$ quartic
superpotential, $h_i$, have a dependence on $t$, which is a coordinate
on the moduli space of the theory. This is natural, because the
quartic superpotential is computed in the background of the VEVs for
the massive fields,  $\langle \Phi_i\rangle\sim t$. It would be
interesting to understand this running of couplings in  $W(\Phi_i)$ 
better, both in field theory and in the 
gravity dual. It is certainly related to the position dependence of
the flux on the dual geometry. 

We also saw that the nonconformal theories obtained by adding
fractional branes are quite interesting.  We saw that the duality
cascade in these theories seems to take theories with 
$\sum_i h_i^{-1}\neq 0$ onto a theory with $\sum_i h_i^{-1}=0$ in the
IR. Correspondingly, the moduli space of scalars is growing an extra
complex dimension at the end of the cascade. Quantum corrections to
the superpotential prevented us from making 
a decisive demonstration of this, however.  We also saw that quantum
corrections lead to interesting deformations of the moduli spaces of
the field theory.

It is important to match the gauge theory and gravity descriptions in
these cases.  Better knowledge of the gravity solution should shed
light on the nature of the corrections to the superpotential. The
metric structure of the solution should be very close to that of the
base of the generalized conifold. Then the solutions for the flux (in
particular the 5--form), will be very important for computing the
moduli space of a D--brane probe. The differences between the moduli
space geometry and that of the original generalized conifold will
reflect the corrections to the field theory superpotential.  

These corrections would be important for settling the issue of whether
the theories do in fact cascade onto theories with 
$\sum_i h_i^{-1}=0$. It would be interesting to elaborate upon the
corresponding behavior of the 3--form flux in the gravity dual. 

\bigskip\bigskip
\noindent
{\large {\bf Acknowledgments}}

We would like to thank 
Jaume Gomis, Ken Intriligator, Clifford Johnson, Rob Leigh, Nick Jones, 
Sheldon Katz, Joe Polchinski, Christian R\"omelsberger, Matt Strassler, 
Johannes Walcher and Nick Warner for useful discussions. 
N.H.\ would especially like to thank Chris Herzog for numerous
enjoyable and helpful discussions during his time at the KITP. 
This work was initiated and largely completed whilst N.H.\ was a 
graduate fellow at the KITP in Santa Barbara. The KITP Graduate Fellowship 
Program is a remarkable scientific opportunity and a whole 
lot of fun.
N.H.\ would also like to thank Brookie Williams
for asking how his day was. 
The work of R.C.\ was supported in
part by the U.S. Department of 
Energy under contract DE-FG02-91ER40677. The work of N.H.\ was
supported in part by a Fletcher Jones Graduate Fellowship from USC
and the National Science Foundation under grant
number PHY99-07949.

\bigskip
\bigskip

\providecommand{\href}[2]{#2}\begingroup\raggedright\endgroup

\end{document}